\documentclass[iop]{emulateapj}

\usepackage{natbib}
\usepackage{amsmath}
\usepackage{url}

\newcommand{\sname}{W0948+0318}

\begin{document}

\title{A Luminous Transient Event in a Sample of WISE-Selected
  Variable AGN}

\author{R.J.~Assef\altaffilmark{1},
  J.L.~Prieto\altaffilmark{1,2},
  D.~Stern\altaffilmark{3},
  R.M.~Cutri\altaffilmark{4},
  P.R.M.~Eisenhardt\altaffilmark{3},
  M.J.~Graham\altaffilmark{5},
  H.D.~Jun\altaffilmark{6},
  A.~Rest\altaffilmark{7,8},
  H.A.~Flewelling\altaffilmark{9},
  N.~Kaiser\altaffilmark{10},
  R.-P.~Kudritzki\altaffilmark{9,11},
  C.~Waters\altaffilmark{9}}

\altaffiltext{1}{N\'ucleo de Astronom\'ia de la Facultad de
  Ingenier\'ia y Ciencias, Universidad Diego Portales, Av. Ej\'ercito
  Libertador 441, Santiago, Chile. Email: roberto.assef@mail.udp.cl}

\altaffiltext{2}{Millennium Institute of Astrophysics, Santiago, Chile}

\altaffiltext{3}{Jet Propulsion Laboratory, California Institute of
  Technology, 4800 Oak Grove Drive, Pasadena, CA 91109, USA}

\altaffiltext{4}{IPAC, Mail Code 100-22, California Institute of
  Technology, 1200 E. California Blvd, Pasadena, CA 91125, USA}

\altaffiltext{5}{Cahill Center for Astronomy and Astrophysics,
  California Institute of Technology, 1216 E. California Blvd.,
  Pasadena, CA 91125, USA}

\altaffiltext{6}{School of Physics, Korea Institute for Advanced Study, 85
  Hoegiro, Dongdaemun-gu, Seoul 02455, Republic of Korea}

\altaffiltext{7}{Space Telescope Science Institute, 3700 San Martin
  Drive, Baltimore, MD 21218, USA} 

\altaffiltext{8}{Department of Physics and Astronomy, The Johns
  Hopkins University, 3400 North Charles Street, Baltimore, MD 21218,
  USA}

\altaffiltext{9}{Institute for Astronomy, University of Hawaii at
  Manoa, 2680 Woodlawn Drive, Honolulu HI 96822, USA}

\altaffiltext{10}{D\'epartement de Physique, \'Ecole Normale
  Sup\'erieure 24 Rue Lhomond 75231 Paris CEDEX 05, France}

\altaffiltext{11}{University Observatory Munich,
  Ludwig-Maximilians-University Munich, Scheinerstr. 1, D-81679
  Munich, Germany}

\begin{abstract}
Recently \citet{assef18} presented two catalogs of AGN candidates over
30,093~deg$^2$ selected from the Wide-field Infrared Survey Explorer
(WISE) observations. From their most reliable sample, \citet{assef18}
identified 45 AGN candidates with the highest variability levels in
the AllWISE catalog, but that are not blazars. Here we present new
spectroscopic observations of some of these targets to further
constrain their nature. We also study their optical lightcurves using
observations from CRTS, and find that only seven show significant
optical variability, and that five of those seven are
spectroscopically classified as AGN. In one of them, WISEA
J094806.56+031801.7 (\sname), we identify a transient event in the
CRTS lightcurve. We present a detailed analysis of this transient, and
characterize it through its CRTS lightcurve and its multi-wavelength
spectral energy distribution obtained from GALEX, Pan-STARRS and WISE
observations. We find that the most likely source of the transient is
a super-luminous supernova (SLSN) in \sname. We estimate the total
radiated energy to be $E=1.6\pm 0.3\times 10^{52}~\rm erg$, making it
one of the most energetic SLSN observed. Based on the lack of change
in mid-IR color throughout and after the transient event, we speculate
that the location of the SLSN is within the torus of the AGN. We
identify 9 possible analogs to \sname\ based on their WISE
lightcurves. None show optically detected transients and hence suggest
significant dust obscuration. Finally, we estimate a rate of $>2\times
10^{-7}~\rm yr^{-1}$ per AGN for these transients under the
conservative assumption that none of the identified analogs have a
common origin with the transient in \sname.
\end{abstract}


\section{Introduction}

Active Galactic Nuclei (AGNs) have long been recognized as
intrinsically variable sources. This variability has been used not
only as a tool to identify AGN \citep[see, e.g.,][and references
  therein]{padovani17}, but also to study physical characteristics of
AGN themselves, such as their black hole masses
\citep{peterson04,vestergaard06} and the structure of their broad-line
region \citep[BLR;]{denney09,pancoast11,denney12}. Although the
mechanism is yet to be understood, the brightness of their accretion
disks changes stochastically as a function of time. This variability
has been best studied at optical wavelengths, where it has been shown
to follow a well defined structure function whose amplitude changes as
a function of wavelength, black hole mass and accretion rate
\citep[see, e.g.,][]{vandenberk04,macleod10}. For most quasars, the
variability can be well characterized by a damped random walk (DRW)
process at intermediate and long timescales
\citep{kelly09,kozlowski10c,macleod10,zu11}. AGN variability has also
been well studied at X-ray wavelengths \citep[e.g.,][and references
  therein]{padovani17}, where emission is dominated by the
corona. Studies have shown that variability at these wavelengths
qualitatively share some of the characteristics found in the optical,
although with significantly shorter timescales for similar variability
amplitudes \citep[e.g.,][]{lawrence87} and a break in the power
spectral density distribution that depends on black hole mass
\citep[e.g.,][]{mchardy06}.

In the mid-IR, however, there have not been many surveys that can
address variability. At these wavelengths, the emission is dominated
by the so-called dust torus \citep[see, e.g.,][]{nenkova08}, which
simply reprocesses the emission from the accretion disk into the
mid-IR. The most notable studies of AGN variability at these
wavelengths have come from the {\it{Spitzer Space Telescope}} Deep,
Wide-Field Survey \citep[SDWFS;][]{ashby09}, and its follow-up, the
Decadal IRAC Bo\"otes Survey (DIBS). Based on four and five epochs of
{\it{Spitzer}} observations, respectively, \citet{kozlowski10} and
\citet{kozlowski16} found that the variability in the mid-IR is also
well characterized by a single structure function with a similar slope
as found at optical wavelengths \citep{caplar17}.  Additionally, from
these observations \citet{kozlowski10b} identified a self-obscured
luminous supernova at $z=0.19$ that was only detected in the mid-IR,
without a clear optical transient counterpart.

A new window into mid-IR variability has been opened by the Wide-field
Infrared Survey Explorer \citep{wright10}, which scanned the entire
sky in four mid-IR bands, centered at 3.4, 4.6, 12 and 22$\mu$m, and
referred to as W1--W4, respectively. Due to the survey scanning
strategy, WISE provided variability on short and long timescales for
the entire sky, with the number of observations being a strong
function of ecliptic latitude. The WISE all-sky survey was conducted
between January and September of 2010, at which point the cryogen was
exhausted. The NEOWISE survey \citep{mainzer11} continued the
observations in the two shortest wavelength bands, W1 and W2, until
February 2011. After a nearly two-year hiatus, the survey then resumed
as the NEOWISE Reactivation mission \citep[NEOWISE-R;][]{mainzer14} in
December 2013 to continue surveying the sky. The AllWISE data
release\footnote{http://wise2.ipac.caltech.edu/docs/release/allwise/expsup/}
made available all the processed observations obtained through 2011,
as well as a catalog of all detected sources.

\citet{assef18} recently presented a nearly all-sky catalog of AGN
from the WISE mission, selected from their W1 and W2 photometry using
a variation of the selection criteria developed by \citet{stern12} and
\citet{assef13}. Using the AllWISE data release of the WISE mission,
\citet{assef18} presented two catalogs: one based on a 90\%
reliability selection criteria (i.e., where 90\% of the selected
objects are expected to be bona-fide AGN) referred to as R90, with
approximately 4.5 million AGN candidates, and one based on a 75\%
completeness criteria (i.e., where 75\% of AGN detected by WISE are
recovered), referred to as C75, with approximately 21 million AGN
candidates \citep[see][for details]{assef18}. Due to its survey
design, WISE naturally obtains variability information on two
timescales: one of approximately 3~hrs, equal to twice the orbital
period of the spacecraft, and another one regulated by the orbital
motion of the Earth, which corresponds to 0.5~years in the ecliptic,
and becomes significantly smaller near the ecliptic poles, achieving
visibility during the full survey duration at the latter. For the rest
of this article, we refer to each group of observations separated by
the $\sim$0.5~year timescale as the epochs of the WISE observations.

\citet{assef18} noted that 687 of the $\sim$4.5 million AGN candidates
in the R90 catalog (0.015\%) were classified as highly variable (i.e.,
variability flag {\tt{var\_flg}}$=$9) in the AllWISE database in both
W1 and W2. In brief, AllWISE characterizes the probability that the
flux of any given source in a specific band is not constant with time
in individual exposures, based on the flux uncertainties and the
temporal correlation with the flux at different bands. A variability
flag of {\tt{var\_flg}}$=$9 (0) indicates the highest (lowest)
probability of variability. Sources with {\tt{var\_flg}} of 8 or 9 are
most likely truly variable in a band, while sources with
{\tt{var\_flg}}$\leq$5 are most likely not real variables. As
expected, \citet{assef18} found the majority of these sources to be
likely blazars, as of the 207 sources located within the FIRST survey
\citep{becker95} footprint, 162 (78\%) have well detected radio
emission in the FIRST
catalog\footnote{http://sundog.stsci.edu/cgi-bin/searchfirst} within
5\arcsec\ from the WISE position\footnote{One additional source, WISEA
  J151215.73+020316.9, is associated to two radio sources,
  9\arcsec\ and 13\arcsec\ away, and has been associated to the
  $\Gamma$-ray source 3FGL J1512.2+0202, making it most likely a
  blazar. For compatibility with the list of targets of
  \citet{assef18}, we keep it in the list of radio-undetected sources
  discussed in the next sections.}. For the other 45 sources covered
by FIRST, however, the underlying mechanism for their variability
remains unclear.  One of these sources, WISEA J142846.71+172353.1, was
identified as a changing look AGN by \citet{assef18} by comparing
spectroscopy from the Sloan Digital Sky Survey (SDSS) DR13
\citep{albareti17} obtained in 2008 ($\sim$2 years before the WISE
survey observations) with dedicated follow-up spectroscopic
observations obtained in 2017. Due to the large change in flux
observed in the WISE bands, \citet{assef18} were able to determine the
most likely source of the dramatic AGN variability to be a drop in the
accretion rate rather than a change in obscuration \citep[see][for
  detailed discussion of the physics behind similar
  objects]{ross18,stern18}.

In this article we focus on these 45 highly variable mid-IR AGN
identified by \citet{assef18} that are within the FIRST survey
footprint but are undetected at radio wavelengths. In
\S\ref{sec:var_wise_AGN} we discuss the spectroscopic classification
of these sources, presenting new observations to update the discussion
of \citet{assef18}. In this section we also study the optical
lightcurves of these sources, and identify one source, WISEA
J094806.56+031801.7 (\sname), that shows a unique lightcurve. In
\S\ref{sec:obs_w0948} we present additional observations of \sname,
and in \S\ref{sec:analysis_w0948} we present a detailed discussion of
its lightcurve. In \S\ref{sec:source_w0948} we present a discussion of
the possible sources for the observed transient, and in
\S\ref{sec:analogs} we present some possible analogs to this transient
that have very similar mid-IR lightcurves but show no significant
optical variability. Throughout the paper, magnitudes are presented in
their natural system unless stated otherwise, namely Vega for $V_{\rm
  CSS}$ and the WISE bands, and AB for $gri$ and the GALEX bands. We
assume a flat $\Lambda$CDM cosmology with $\Omega_{\rm M} = 0.286$ and
$H_0=69.3~\rm km~\rm s^{-1}~\rm Mpc^{-1}$.

\section{Highly Variable Radio-Undetected WISE AGN}\label{sec:var_wise_AGN}

In this section we focus on the 45 highly variable AGN candidates
identified by \citet{assef18} from their 90\% reliability WISE AGN
catalog (R90) that are within the FIRST survey footprint but are
undetected at radio wavelengths. 

\subsection{Spectroscopic Classification}\label{ssec:var_agn_spec_class}

\citet{assef18} studied the spectroscopic classification of 34 of
these 45 sources, with 32 of them taken from either SDSS or
SIMBAD\footnote{\url{http://simbad.u-strasbg.fr/simbad/}}, and two of
them determined from recent observations carried out using the DBSP
instrument at the Palomar Observatory 200-inch telescope
\citep[see][for details]{assef18}. They found that 31 of the 34
objects were consistent with an AGN classification based on their
optical spectra, and that the remaining three objects were stellar
contaminants (consistent with the expected 90\% reliability of the
sample). Two of the Galactic interlopers were classified as carbon
stars. The AGN have redshifts between 0.0365 and 0.5550 and have a
diversity of spectral features.

Of these 31 objects consistent with an AGN spectroscopic
classification, 18 were classified as QSOs, one was classified as a
Seyfert 1 galaxy, one as a red type 1 AGN, and one as a type 2
AGN. Six other objects were classified as Galaxy AGN by SDSS, and
inspection of their spectra indicates that either they have broad
bases to their H$\alpha$ emission lines but only show narrow, if any,
H$\beta$ emission, or that they have line ratios indicative of
obscured AGN activity. The remaining four objects were classified as
galaxies, but \citet{assef18} find, upon inspection of their spectra,
that all show hints of broad bases to their H$\alpha$ emission lines,
suggesting an important AGN contribution. Hence, of the 31 highly
variable AGN with spectroscopic classifications, 19 (61\%) are
unobscured AGN, and the remaining 12 (39\%) show some degree of likely
obscuration by either being classified as red type 1,
intermediate-type, or type 2 AGN.

To provide a more complete spectroscopic analysis, we have obtained
spectroscopic observations of an additional four targets from the
sample of 45. We also obtained spectroscopic observations of WISEA
J150954.94+203619.6 for which, although discussed by \citet{assef18},
only a photometric classification of ``Possible AGN'' and a
photometric redshift of 0.41492 were reported by SIMBAD from the work
of \citet{oyaizu08} and \citet{szabo11}. We obtained the spectra using
the DBSP optical spectrograph at the Palomar Observatory 200 inch
telescope, on the nights of UT 2016 February 6, UT 2017 April 22, UT
2017 August 6 and UT 2018 January 6. We used the same instrumental
setup as described in \citet{assef18}, namely a slit with a
1.5\arcsec\ width, the D55 dichroic, the 600 lines/mm grating
(4000\AA\ blaze) on the blue arm, and the 316 lines/mm grating
(7500\AA\ blaze) on the red arm. Reductions were carried out in a
standard manner using IRAF\footnote{http://iraf.noao.edu}.

\begin{figure}
  \begin{center}
    \plotone{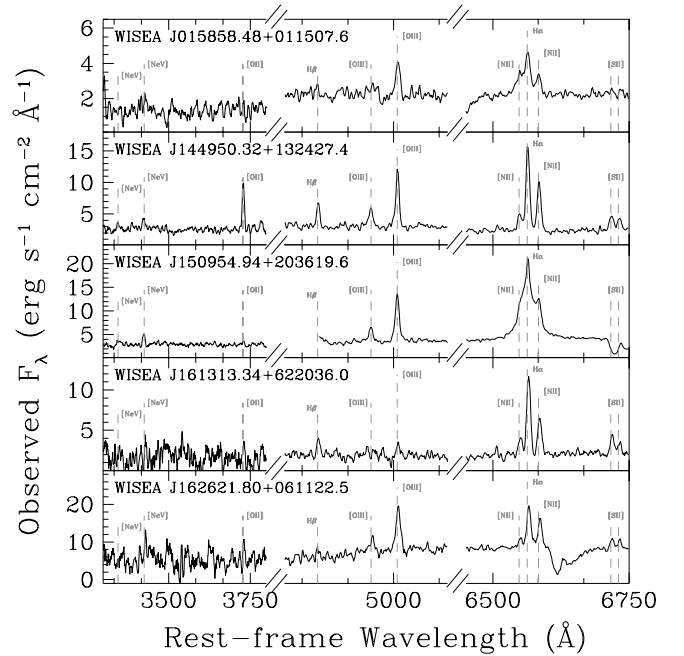}
    \caption{Optical spectra of five radio-undetected, highly variable
      AGN candidates. All spectra were obtained with the DBSP
      instrument at the Palomar Observatory 200-inch telescope. The
      gap in the spectrum of WISEA J150954.94+203619.6 is due to the
      dichroic used in the observations. Note that the x-axis is split
      into three different wavelength ranges for clarity.}
    \label{fg:p200_specs}
  \end{center}
\end{figure}

The spectra of the five targets are shown in Figure
\ref{fg:p200_specs}. Two of the sources, WISEA J015858.48+011507.6
($z=0.184$) and WISEA J150954.94+203619.6 ($z=0.131$) have broad
H$\alpha$ emission lines, consistent with an AGN classification. The
latter has a spectroscopic redshift well below the photometric
redshift of 0.41492. Two other sources, WISEA J144950.32+132427.4
($z=0.198$) and WISEA J162621.80+061122.5 ($z=0.149$) have high
[O\,{\sc iii}] to H$\beta$ ratios and detections of [Ne\,{\sc v}]
emission, consistent with an obscured AGN classification. The other
source, WISE J161313.34+622036.0 ($z=0.262$), has only narrow emission
lines that do not suggest an AGN classification. This implies the
source is likely a contaminant to the R90 sample, as a certain number
of low redshift, strongly star-forming galaxies are expected to
contaminate the sample \citep[see][for details]{assef18}. An updated
version of Table 4 of \citet{assef18} is provided in Table
\ref{tab:var_rq} including the spectroscopic redshifts and
classifications of these five targets as well all the other highly
variable AGN with and without spectroscopic information.

\begin{deluxetable*}{l c c l l}

  \tablecaption{Properties of Radio-Quiet, Highly
    Variable WISE AGN\label{tab:var_rq}}

  \tablehead{
    \colhead{WISE ID}&
    \colhead{Redshift}&
    \colhead{CRTS $\chi^2_{\nu}$}&
    \colhead{Classification}&
    \colhead{Ref}\\
    \colhead{(WISEA)}&
    \colhead{}&
    \colhead{}&
    \colhead{}&
    \colhead{}
  }

  \tabletypesize{\small}
  \tablewidth{0pt}
  \tablecolumns{5}

  \startdata
  J000011.72+052317.4  & 0.0400     &     2.4 & Seyfert 1    & SIMBAD\\
  J014004.69--094230.4 & 0.1461     &     2.3 & QSO          & SDSS\\
  J015858.48+011507.6  & 0.184\phn  &     2.3 & AGN          & P200/DBSP\\
  J090546.35+202438.2  &\nodata     &    20.5 & Carbon Star  & SIMBAD\\
  J091225.00+061014.8  & 0.1453     &     0.6 & Galaxy\tablenotemark{$\dagger$} & SDSS\\
  J094806.56+031801.7  & 0.2073     &     4.9 & QSO          & SDSS\\
  J095824.97+103402.4  & 0.0417     &     1.3 & Galaxy AGN   & SDSS\\
  J100933.13+232255.7  & 0.0719     &     0.4 & Galaxy AGN   & SDSS\\
  J101536.17+221048.9  & \nodata    &     2.2 & \nodata      & \nodata\\
  J104241.08+520012.8  & 0.1365     &     2.1 & QSO          & SDSS\\
  J112537.83+212042.2  & 0.0894     &     2.0 & QSO          & SDSS\\
  J130155.84+083631.7  &\nodata     &    16.4 & Carbon Star  & SIMBAD\\
  J130716.98+450645.3  & 0.0843     &     0.9 & QSO          & SDSS\\
  J130819.11+434525.6  & 0.0365     &     0.4 & Galaxy AGN   & SDSS\\
  J140033.66+154432.1  & 0.2152     &     2.2 & QSO          & SDSS\\
  J141053.43+091027.0  & 0.1781     &     7.1 & QSO          & SDSS\\
  J141105.45+294211.8  & 0.0724     &     0.5 & QSO          & SDSS\\
  J141758.60+091609.7  & 0.1389     &     3.6 & QSO          & SDSS\\
  J142747.45+165206.0  & 0.1435     &     0.7 & QSO          & SDSS\\
  J142846.71+172353.1  & 0.1040     &     0.7 & QSO          & SDSS\\
  J143457.66-051038.9  & \nodata    &     0.3 & \nodata      & \nodata\\
  J144039.30+612748.1  & 0.0811     & \nodata & QSO          & SDSS\\
  J144131.81+321612.9  & 0.1993     &     0.7 & QSO          & SDSS\\
  J144439.59+351304.7  & 0.0790     &     0.4 & Galaxy\tablenotemark{$\dagger$}  & SDSS\\
  J144510.14+304957.1  & 0.2754     &     6.1 & QSO          & SDSS\\
  J144603.98--013203.4 & 0.0840     &     0.3 & Galaxy AGN   & SDSS\\
  J144950.32+132427.4  & 0.198\phn  &     0.9 & AGN          & P200/DBSP\\
  J145222.03+255152.0  & 0.1204     &     0.4 & QSO          & SDSS\\
  J150842.68+212132.1  & \nodata    &     1.2 & \nodata      & \nodata\\
  J150954.94+203619.6  & 0.131\phn  &     0.6 & AGN          & P200/DBSP\\
  J151215.73+020316.9  & 0.2199     &     9.7 & Galaxy AGN\tablenotemark{$\ddagger$}   & SDSS\\
  J151444.52+364237.9  & 0.1944     &     1.0 & QSO          & SDSS\\
  J151518.56+312937.5  & 0.1036     &     0.5 & QSO          & SDSS\\
  J154324.98+545652.0  & \nodata    &     1.2 & \nodata      & \nodata\\
  J155223.29+323455.0  & 0.1277     &     1.5 & Galaxy\tablenotemark{$\dagger$} & SDSS\\
  J161313.34+622036.0  & 0.262\phn  &     1.4 & Galaxy       & P200/DBSP\\
  J161846.36+510035.1  & 0.3189     &     1.9 & QSO          & SDSS\\
  J162140.25+390105.1  & 0.0642     &     0.6 & Galaxy AGN   & SDSS\\
  J162621.80+061122.5  & 0.149\phn  &     1.0 & AGN          & P200/DBSP\\
  J163214.82+634854.5  & \nodata    &     1.5 & \nodata      & \nodata\\
  J163518.38+580854.6  & \nodata    &     1.2 & Star         & SIMBAD\\
  J165336.08+594347.1  & \nodata    &     0.9 & \nodata & \nodata\\
  J165932.68+470448.5  & \nodata    &     0.8 & \nodata & \nodata\\
  J170741.24+523714.4  & \nodata    &     0.8 & \nodata & \nodata\\
  J213604.22--050152.0 & 0.1284     &     0.8 & Galaxy\tablenotemark{$\dagger$} & SIMBAD\\
  \enddata
  

  \tablenotetext{$\dagger$}{Although the object is classified as a
    galaxy in SDSS or SIMBAD, the H$\alpha$ emission line shows a
    broad base suggesting the presence of an AGN.}

  \tablenotetext{$\ddagger$}{Most likely a blazar due to its association
  with the $\Gamma$-ray source 3FGL J1512.2+0202 and large scale radio
  emission.}

\end{deluxetable*}

\subsection{Lightcurves}\label{ssec:var_agn_lightcurves}

To further characterize these 45 highly variable WISE AGN candidates,
we cross-matched our sample with optical photometry available from the
Catalina Real-Time Transient Survey \citep[CRTS;][]{drake09} DR2 data
release\footnote{\url{http://nesssi.cacr.caltech.edu/DataRelease/}}. CRTS
DR2 provides optical lightcurves in the $V_{\rm CSS}$ band for objects
with declinations in the range $-75~\rm deg\lesssim\delta\lesssim
65~\rm deg$ and farther than 10--15~deg from the Galactic Plane with
$V_{\rm CSS}$-band magnitudes between 11.5 and 21.5. We obtained CRTS
optical lightcurves for 44 of the 45 highly variable, radio-undetected
WISE AGN candidates. The source without a CRTS counterpart, WISEA
J144039.30+612748.1, is outside the footprint of the CRTS survey.

\begin{figure}
  \begin{center}
    \plotone{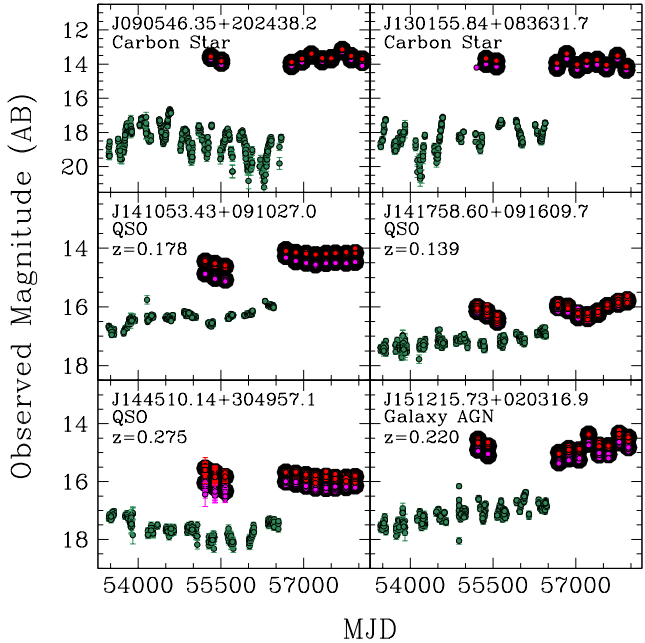}
    \caption{Lightcurves of six of the seven radio-undetected, highly
      mid-IR variable AGN candidates with high optical variability
      (see text for details). The other target is shown in
      Fig. \ref{fg:lightcurve}. The figure shows the optical CRTS
      $V_{\rm CSS}$-band (green) and WISE W1 (magenta) and W2 (red)
      bands. For W1 and W2, the small circles show the individual
      frame photometry in the AllWISE and NEOWISE-R surveys. The large
      black circles show the median of each epoch. To minimize the
      range of magnitudes between bands, all magnitudes are shown in
      the AB system. To convert W1 (W2) from Vega to AB, we add
      2.68~mag (3.32~mag).}
    \label{fg:lcs_high_var}
  \end{center}
\end{figure}

\begin{figure}
  \begin{center}
    \plotone{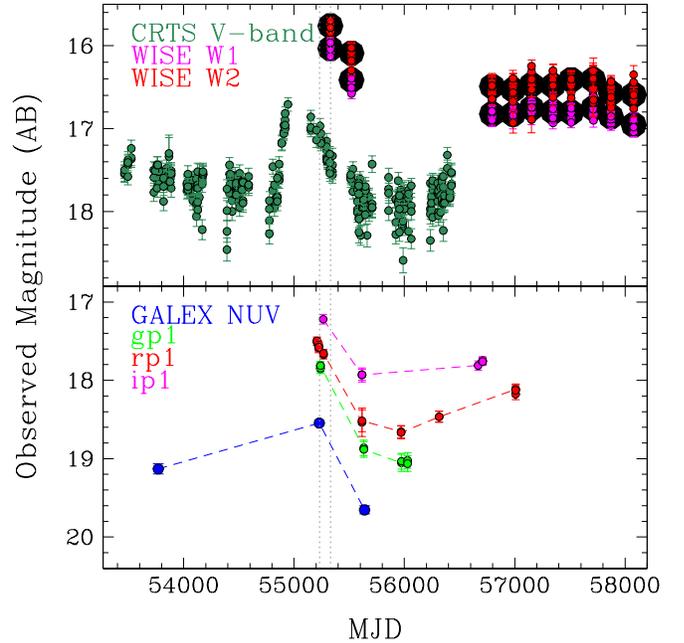}
    \caption{Lightcurve of WISE J094806.56+031801.8 (\sname). Symbols
      in the top panel have the same definition as in
      Fig. \ref{fg:lcs_high_var}. The bottom panel shows the GALEX NUV
      band (blue) and Pan-STARRS $g$ (green), $r$ (red) and $i$
      (magenta). All magnitudes are in the AB system. The vertical
      gray dotted lines show the dates at which we model the SED of
      the transient in \S\ref{ssec:transient_sed}.}
    \label{fg:lightcurve}
  \end{center}
\end{figure}

Inspection of the optical lightcurves reveals that while some objects
show strong optical variability, the majority of objects are
relatively quiescent, despite being highly variable in the mid-IR. We
quantify the level of variability of sources by comparing how much the
lightcurve deviates from a constant flux. Specifically, we fit a
constant flux to the $V_{\rm CSS}$ lightcurve and measure
$\chi^2_{\nu}$, the $\chi^2$ per degree of freedom, of the fit. We
eliminate the 5\% most discrepant points from each lightcurve to
ensure that high $\chi^2_{\nu}$ values are driven by real variability
and not outliers. These values are shown for all 44 lightcurves in
Table \ref{tab:var_rq}. Note that 22 (50\%) have $\chi^2_{\nu}<1$, and
30 (67\%) have $\chi^2_{\nu}<2$, confirming the majority of targets
have optical lightcurves that are not strongly variable. There are,
however, seven targets, that have $\chi^2_{\nu}>3$, implying stronger
optical variability. Figures \ref{fg:lcs_high_var} and
\ref{fg:lightcurve} show the optical and mid-IR lightcurves of these
seven targets. The lightcurves for the rest of the sample are
presented in \S\ref{sec:analogs} and Appendix
\ref{app:lightcurves_others}. The CRTS lightcurves have been corrected
for foreground reddening using $V$-band extinction determined from the
extinction maps of \citet{schafly11} queried through the NASA/IPAC
Extragalactic Database (NED) Coordinate \& Extinction
Calculator\footnote{\url{https://ned.ipac.caltech.edu/forms/calculator.html}}. To
extend their mid-IR lightcurves we have also included the observations
of the NEOWISE-R mission \citep{mainzer14} from the 2018 Data
Release\footnote{http://wise2.ipac.caltech.edu/docs/release/neowise/}.

\sname\ is unlike any of the other sources presented above, with very
strong epoch-to-epoch variability during the AllWISE observations, but
shows little to no variability during the NEOWISE-R
observations. Concurrent to the AllWISE observations, CRTS shows a
very strong optical transient event whose lightcurve decay
qualitatively matches the behavior observed in the WISE bands. While
other highly variable radio-undetected AGN candidates have similar
WISE lightcurves (see \S\ref{sec:analogs} for details), none show
significant optical variability, and of all targets with high optical
variability, \sname\ is the only one whose optical lightcurve is
dominated by a single transient event.

Recently, \citet{graham17} studied AGN lightcurves in the CRTS data
and independently identified \sname\ as one of 51 objects whose light
curves strongly deviated from a damped random walk (DRW) model. They
concluded that AGN flaring events are incompatible with the stochastic
variability properties of AGN, and instead proposed a number of
different mechanisms for these optical transient events. Specifically,
they discussed these sources in the context of microlensing events by
stars in the Milky Way, super-luminous supernovae, tidal disruption
events (TDEs), and black hole binary mergers. While the data available
did not allow to differentiate between most of these scenarios, they
were able to put constraints on the possible microlensing nature of
these sources, and for the specific case of \sname, \citet{graham17}
find that the microlensing explanation is unlikely given the asymmetry
of the optical lightcurve (see Fig. \ref{fg:lightcurve}). Here, we can
further rule out this scenario since the transient event is also
present in the WISE bands. While the accretion disk can be of
comparable size to the Einstein ring of a stellar lens, the observed
emission in the W1 and W2 bands is dominated by dust in the torus,
which is at significantly larger physical scales.

In the following sections we use additional observations to constrain
the nature of the transient event in \sname.

\section{Additional Observations of \sname}\label{sec:obs_w0948}

\subsection{GALEX Photometry}\label{ssec:galex_phot}

The Galaxy Evolution Explorer \citep[GALEX;][]{martin05}, through its
DR7 data release, provides three epochs of NUV photometry for \sname,
as well as one epoch of photometry in the FUV band. The earliest epoch
provided comes from the coaddition of two exposures obtained
approximately one month apart from each other, on UT 2006-02-03 and
2006-03-16, well before the optical transient. The combined exposure
time is 217~s in each band. For simplicity we assign the epoch of this
observation to the date of the first exposure. This does not impact
any of the results presented in this article. The second epoch
provided also comes from the coaddition of two exposures, but this
time taken only a few hours apart from each other on UT
2010-02-02. This observation was obtained only in the NUV band, with a
total exposure time of 2119~s. The third and final epoch comes from a
single NUV exposure of 1648.15~s obtained on UT 2012-03-06.

The target is detected in the NUV band in all three visits, with
reported respective AB magnitudes of 19.410$\pm$0.065,
18.825$\pm$0.021 and 19.933$\pm$0.052. The target was detected in the
only FUV observation obtained, with a magnitude of
19.472$\pm$0.105. The NUV lightcurve is presented in the bottom panel
of Figure \ref{fg:lightcurve}. By chance, the second epoch of GALEX
observations was obtained shortly after the maximum of the transient
event, and the third epoch was obtained shortly after the transient
had faded away. The first epoch, instead, was obtained significantly
before the transient event, towards the beginning of the CRTS
observations. For the spectral energy distribution (SED) analysis
presented in \S\ref{sec:analysis_w0948}, we correct the GALEX
photometry for foreground reddening using the coefficients of
\citet{yuan13} to extrapolate the $u$-band reddening reported by NED
from the \citet{schafly11} extinction maps to the NUV band. For FUV we
assume the same extinction as for NUV due to the uncertainties in the
coefficient reported by \citet{yuan13}, and add to the uncertainty in
FUV a factor based on the ranges of coefficients they considered.

\subsection{Pan-STARRS Photometry}\label{ssec:panstarrs_phot}

Starting in 2010, the Panoramic Survey Telescope and Rapid Response
System \citep[Pan-STARRS;][]{chambers16,flewelling16} has used a 1.8~m
telescope to repeatedly image the northern sky in the $grizy$
bands. During its first years of operations, the survey observed
\sname\ in the $g$, $r$ and $i$ bands during the transient event as
well as after it had disappeared. Individual observations in the $g$
band were obtained with exposure times of either 52 or 43~s, with
exposure times ranging between 30 and 60~s in the $r$ band, and with
exposure times of either 32 or 45~s in the $i$ band. The basic data
processing of these individual PS1 images is performed by the PS1 IPP
system \citep{magnier16,waters16}. Downstream processing is then
performed with the {\tt{photpipe}} pipeline which has been used for
various time-domain surveys like SuperMACHO, ESSENCE, and the PS1
Medium Deep Field survey \citep{rest05,rest14}. Images in the $z$ and
$y$ bands were only obtained after the transient had faded away. We
measured the flux of the source for every epoch of Pan-STARRS using
4\arcsec\ diameter apertures. We obtained the zero-point of the
photometry by comparing the flux of all stars in the field against
their magnitudes reported by the Pan-STARRS DR1 for the stacked frames
(photometry for individual exposures is not available in DR1). We
correct the photometry for foreground reddening using the extinction
maps of \citet{schafly11} queried through NED. The respective
lightcurves are shown in the bottom panel of Figure
\ref{fg:lightcurve}.

In the coadded frames released as part of the PanSTARRS DR1,
\sname\ is marked as extended. However, in the $g$, $r$ and $i$ bands
the difference between the profile fitting magnitude and the
Kron-radius magnitude is $\sim$0.05~mag, indicating the majority of
the flux is consistent with a point source and hence that the source
is only marginally resolved in these bands.

\subsection{SDSS Photometry and Spectroscopy}\label{ssec:sdss_spec}

The Sloan Digital Sky Survey (SDSS) DR13 reports photometry for
\sname\ obtained before the transient event on MJD 51960,
approximately 2800~days ($\sim 7.5~\rm years$) before the transient
event. For the analysis of the pre-transient SED presented in the next
section, we specifically use the {\tt{ModelMag}} AB magnitudes of
$u=$18.80$\pm$0.02, $g=$18.50$\pm$0.01, $r=$18.21$\pm$0.01,
$i=$17.87$\pm$ 0.01 and $z=$17.94$\pm$0.02.

\begin{figure}
  \begin{center}
    \plotone{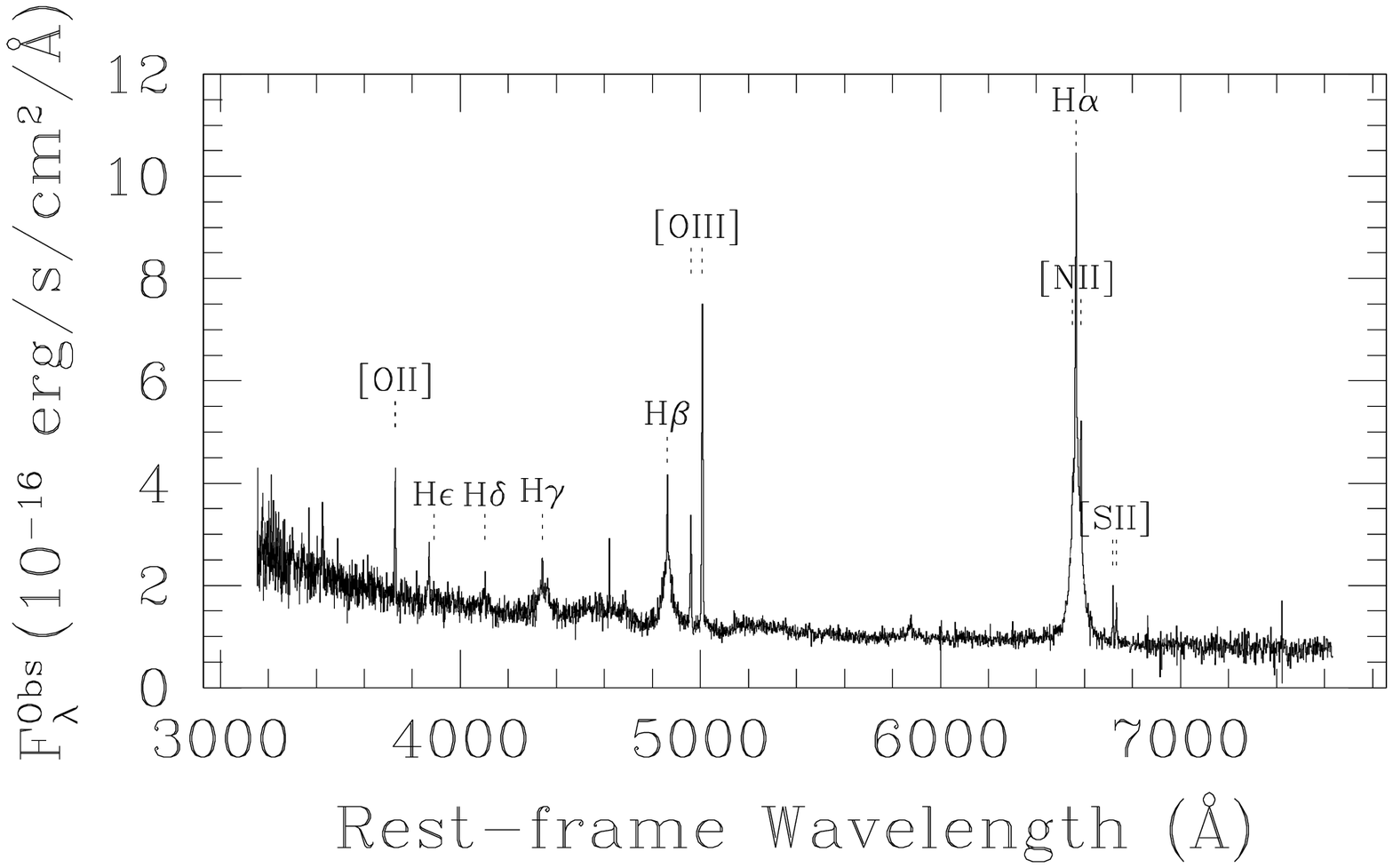}
    \caption{SDSS spectrum of \sname.}
    \label{fg:sdss_spec}
  \end{center}
\end{figure}

\sname\ is spectroscopically identified as a type 1 AGN at a redshift
of $z=0.20728\pm 0.00003$ in the Sloan Digital Sky Survey (SDSS)
DR13. The SDSS optical spectrum is shown in Figure \ref{fg:sdss_spec}
and was obtained in MJD 52266, approximately 2500~days ($\sim 7~\rm
years$) before the event. The hydrogen emission lines are clearly
broadened with respect to the rest of the emission lines, consistent
with the type 1 AGN classification. The continuum luminosity at
5100\AA\ is measured to be $\rm L_{5100} = 8.5\pm 0.5 \times
10^{43}~\rm erg~\rm s^{-1}$. From the photometric SED modeling of
\sname\ before the transient event that will be presented in
\S\ref{ssec:pre_transient_SED_modeling}, we estimate that $41\pm 6$\%
of the emission observed at 5100\AA\ comes from the AGN, with the rest
coming from the host galaxy. Hence, the AGN continuum luminosity at
5100\AA\ is $\rm L_{5100}^{AGN} = 3.5\pm 0.5\times 10^{43}~\rm erg~\rm
s^{-1}$.

The SDSS DR13 reports line-widths for both the H$\alpha$ and H$\beta$
emission lines of $\sigma = 886.5\pm 7.6~\rm km~\rm s^{-1}$, while all
narrow-emission lines, including [N\,{\sc ii}] and [O\,{\sc iii}],
have reported line-widths of $\sigma = 132.8\pm 2.0~\rm km~\rm
s^{-1}$. Using the AGN continuum luminosity at 5100\AA, $\rm
L_{5100}^{AGN}$ and the line dispersion of H$\beta$, we infer a black
hole mass of $M_{\rm BH} = 1.2\pm0.4\times 10^{7}~M_{\odot}$ using the
radius luminosity relation of \citet{bentz09} and the $f$ factor of
\citet{collin06}, following eqn. (4) of \citet{assef11}. Using the
bolometric correction for $\rm L_{5100}^{AGN}$ of \citet{gallagher07},
we infer a bolometric luminosity of $L_{\rm Bol} = 3.6\pm1.6\times
10^{44}~\rm erg~\rm s^{-1}$, which implies an Eddington ratio of
$L_{\rm Bol}/L_{\rm Edd} = 0.25\pm 0.14$. Note, however, that these
could be considered as upper limits, as the bolometric correction used
includes the dust emission, which is just reprocessed light from the
accretion disk. We note that, assuming a Gaussian profile for the
H$\alpha$ and H$\beta$ emission lines, their corresponding FWHM would
be $2088\pm 18~\rm km~\rm s^{-1}$, just above of the usual threshold
of ${\rm{FWHM}} < 2000~\rm km~\rm s^{-1}$ used to classify narrow-line
Seyfert 1s \citep{osterbrock85,goodrich89}. If \sname\ is indeed a
narrow-line Seyfert 1, its Eddington ratio would be towards the
lower-end of the range observed by, e.g., \citet{grupe04}, which goes
from $L_{\rm Bol}/L_{\rm Edd}\sim 0.1$ to somewhat above the Eddington
limit.

\section{Analysis of the Event}\label{sec:analysis_w0948}

\subsection{Pre-Transient SED of \sname}\label{ssec:pre_transient_SED_modeling}

We start by modeling the pre-transient SED of \sname, constructed from
the first epoch of GALEX FUV and NUV observations (see
\S\ref{ssec:galex_phot}), the SDSS photometry reported in
\S\ref{ssec:sdss_spec}, as well as the Two Micron All-Sky Survey
\citep[2MASS;][]{skrutskie06} photometry in the $J$, $H$ and $Ks$
bands reported in the 2MASS All-Sky Point Source Catalog queried
through the Infrared Science Archive (IRSA). The 2MASS observations
were carried out in MJD 51571.31, approximately 1.5~yrs before the
SDSS observations and approximately 9~yrs before the transient event.

We model these bands using the SED templates and algorithm of
\citet{assef10}. In short, we model the SED of the object through a
non-negative linear combination of an empirical AGN template and three
empirical galaxy SED templates, roughly corresponding to a late-type
galaxy (E template), an intermediate spiral (Sbc template) and a local
starburst (Im template). We also fit for the obscuration of the AGN
template \citep[see][for details]{assef10}. We note that no effort was
taken to correct for the intrinsic variability of the AGN between the
different observations. Additionally, we are considering
multi-wavelength photometry obtained through a number of different
methods that likely drive systematic differences between the bands, so
we enforce a minimum flux error of 5\% for each band.

\begin{figure}
  \begin{center}
    \plotone{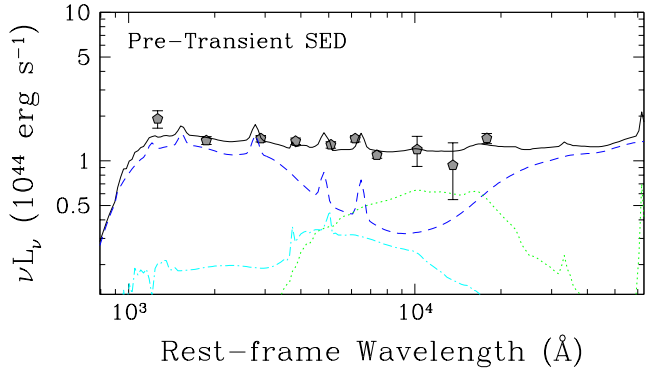}
    \caption{SED of \sname\ before the transient occurred. From
      shortest to longest wavelength, the gray pentagons show the
      monochromatic luminosity of \sname\ in the GALEX FUV and NUV
      bands, the SDSS $u$, $g$, $r$, $i$ and $z$ bands, and the 2MASS
      $J$, $H$ and $Ks$ bands. The solid black line shows the best-fit
      SED model, composed of a mildly reddened AGN (dashed blue line),
      the Sbc galaxy (dotted green line) and the Im galaxy (dot-dashed
      cyan line) templates of \citet{assef10}.}
    \label{fg:AGN_SED}
  \end{center}
\end{figure}

The best-fit SED model is shown in Figure \ref{fg:AGN_SED}, and has
$\chi^2_{\nu} = 2.6$. As expected, the source is dominated by an AGN
component in the UV and optical, while the host galaxy emission
dominates the near-IR. The AGN contributes 76\% of the luminosity in
the wavelength range between 0.1 and 30~$\mu$m. The best-fit SED model
requires a mild reddening of $E(B-V)=0.07$ over the AGN component,
although we note that the FUV band is significantly above the best-fit
SED. This suggests that the true $E(B-V)$ may closer to zero, but that
the reddening is being used to better accommodate the $u$--NUV color
which could be affected by the systematic issues described
above. Furthermore, FUV likely also has a significant contribution
from Ly$\alpha$, which could be stronger than that of the SED template
from \citet{assef10}.

\subsection{Optical Lightcurve}\label{ssec:lightcurve}

Figure \ref{fg:lightcurve} shows the lightcurve of the event. The
optical lightcurve shows a single strong outburst event, starting at
about MJD 54800, with lower amplitude variability over the entire
time-span of the CRTS observations. The transient event is clearly
asymmetric in the optical, brightening in $\sim$200 days, and decaying
over the next $\sim$600 days. The peak of the lightcurve is
unfortunately not well mapped by the CRTS observations, so we assume
it to be at the maximum flux detected, on MJD 54944.23.

Figure \ref{fg:lightcurve} also shows the AllWISE lightcurve, to which
we have added the W1 and W2 observations of the latest data release of
the NEOWISE-R mission. The small red and magenta data points in Figure
\ref{fg:lightcurve} show the individual measurements of each visit,
while the large data points show the median photometry of those visits
grouped by their half year epoch. The error-bars of each individual
visit are somewhat underestimated, so we do not study mid-IR
variability on the shortest timescales. The event was observed by the
WISE survey while fading after its peak magnitude, and shows no
significant variability in the $\sim$6 month timescales during the
first six epochs of the NEOWISE-R mission, while fading in the last
two epochs. The W1--W2 color of \sname\ shows almost no change between
the AllWISE and NEOWISE-R epochs, being consistent with the AGN
classification of \citet{assef18} at all times. The W1--W2 color in
the AllWISE source catalog, which is built on the coaddition of all
the single frames, is 0.934$\pm$0.036. Such value is not available for
NEOWISE-R, but the mean of the single frame measurements is
0.947$\pm$0.118, where the error corresponds to the dispersion of all
individual measurements. Additionally, Figure \ref{fg:lightcurve}
shows the lightcurve of the event in the GALEX NUV band and in the
Pan-STARRS $g$$r$$i$ bands. The transient event is clearly observed in
all four bands presented.

\begin{figure}
  \begin{center}
    \plotone{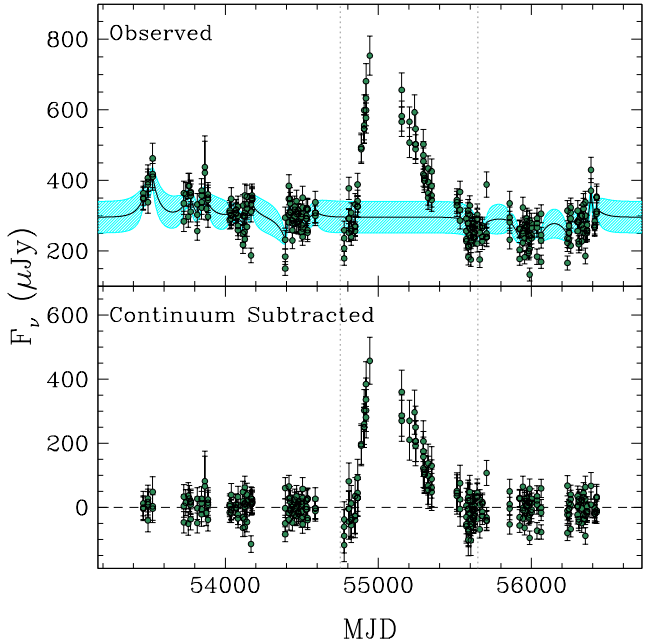}
    \caption{({\it{Top}}) CRTS $V_{\rm CSS}$-band lightcurve of
      \sname. The black solid line shows best-fit DRW model to the
      lightcurve outside the transient event. The cyan region shows
      the dispersion of the DRW lightcurves consistent with the
      data. The vertical dotted lines show the region of the
      lightcurve excised for fitting the DRW model. ({\it{Bottom}})
      $V_{\rm CSS}$-band lightcurve after subtracting the best-fit DRW
      model.}
    \label{fg:javelin_fit_CRTS}
  \end{center}
\end{figure}

In order to isolate the emission of the transient event, we need to
remove the unrelated underlying emission. This is complicated by the
fact that \sname\ is a type 1 AGN, which are intrinsically variable
objects. Indeed, the $V_{\rm CSS}$-band lightcurve in Figure
\ref{fg:lightcurve} shows clear variability, albeit with much lower
amplitude, before and after the transient event. This is also
observable in the GALEX and Pan-STARRS observations although much less
clearly owing to the fewer number of observations. To isolate the
transient in the $V_{\rm CSS}$-band, we model the AGN variability of
\sname\ as a DRW using the code JAVELIN \citep{zu13}. We only use the
emission outside of the transient event, specifically at MJD$<$54750
and MJD$>$55650. The DRW process is modeled by two quantities: the
asymptotic amplitude of the structure function, $SF_{\infty}$, and the
characteristic dampening timescale $\tau$ \citep[see][for
  details]{zu13}. Note that if the region during the transient event
is included in the modeling, the fit does not converge. Figure
\ref{fg:javelin_fit_CRTS} shows the best-fit DRW model and the
continuum subtracted lightcurve of the event. The DRW model of the AGN
lightcurve provides an expected range of dispersion where no data was
available to fit the model. The transient event is well above the
expected AGN variability \citep[also see discussion
  in][]{graham17}. For the subtracted lightcurve we have added in
quadrature the 1$\sigma$ range of the best-fit model to the error of
each data point. Assuming that the magnitude decay is linear with
time, we find a best-fit decay rate for the transient event of
$0.0059\pm 0.0010~\rm mag~\rm day^{-1}$. Assuming the same for the
rise, we get a $\sim$3 times faster brightening rate of $0.019\pm
0.003~\rm mag~\rm day^{-1}$. Note that the uncertainties have been
obtained from a Monte Carlo process taking into account the
uncertainty in the observed fluxes as well as the uncertainty from the
DRW modeling of the underlying emission.

\subsection{Multi-Wavelength Transient SED}\label{ssec:transient_sed}

As little data are available for the Pan-STARRS and GALEX bands, we
cannot model these wavelengths using the DRW approach. Instead we have
simply assumed that the shape of lightcurve is the same as in the CRTS
$V_{\rm CSS}$-band DRW model, scaled to a different mean flux. This is
not unreasonable given that the flux variability in the accretion disk
should be replicated at all wavelengths, with a small time delay
\citep[e.g.,][]{edelson15} and an amplitude that scales with
wavelength \citep[e.g.,][]{macleod10}. For the black hole mass,
Eddington ratio and bolometric luminosity estimated in
\S\ref{ssec:sdss_spec}, assuming the latter corresponds to the
bolometric luminosity of the accretion disk \citep[although see][for
  details]{gallagher07}, the radius of a geometrically thin, optically
thick accretion disk \citep{shakura73} where the thermal radiation
peaks at 5000\AA\ would be 0.26 light-days \citep[eqn. (2)
  of][]{morgan10}. Even though real accretion disks may be $\sim$4
times larger than a thin disk model \citep{morgan10}, the light travel
time within the accretion disk is much shorter than the timescales on
which we are interested, so we ignore this potential concern. To
account for the fact that the amplitude of the variability is
dependent on wavelength, we scale the $SF_{\infty}$ and $\tau$
parameters of the DRW model according to the results of
\citet[][although see \citealt{kozlowski17} who suggests that such
  correlations could be spurious]{macleod10}. Specifically,
$SF_{\infty} \propto \lambda_{\rm RF}^{-0.479}$ and $\tau \propto
\lambda_{\rm RF}^{0.17}$, where $\lambda_{\rm RF}$ is the rest-frame
effective wavelength of each band to scale from the $V_{\rm CSS}$
lightcurve model.

The same approach is not possible for the WISE bands, as the coverage
of the lightcurve is much more sparse, and outside the transient
event, the mid-IR emission is dominated by the dust torus, not by the
accretion disk. Instead, we consider that little variability is
observed during the NEOWISE-R observations, so we simply subtract
their median flux from the AllWISE observations. As the NEOWISE-R
epochs were observed starting $\sim$3.5~yr after the AllWISE epochs,
which is significantly longer than the travel time from the accretion
disk to the torus\footnote{In \S\ref{ssec:sdss_spec} we estimated an
  AGN bolometric luminosity for \sname\ of $L_{\rm Bol} = 3.6\times
  10^{44}~\rm erg~\rm s^{-1}$, although discussed this should be
  considered an upper bound. Using equation (1) of \citet{nenkova08}
  and assuming a dust sublimation temperature of 1500~K, we estimate
  the light travel time to the inner edge of the torus to be $<0.8~\rm
  yr$ as $L_{\rm Bol}$ is an upper bound on the accretion disk
  luminosity.}, we expect the NEOWISE-R emission to be representative
of the typical torus luminosity even if the source of the transient
was within the accretion disk. Furthermore, from the analysis of
\citet{kozlowski16}, we expect $\lesssim 0.2~\rm mag$ variability over
the $\sim$3--4 year timespan between the AllWISE and NEOWISE-R
observations, which is small compared to the transient's amplitude of
$\sim$1~mag. Comparing the two AllWISE epochs after the subtraction,
we find linear magnitude decay rates of $0.0050\pm 0.0013$ and
$0.0045\pm 0.0016~\rm mag~\rm day^{-1}$ respectively in the W1 and W2
bands, similar to the $0.0059\pm 0.0010~\rm mag~\rm day^{-1}$ found
for the $V_{\rm CSS}$-band.

\begin{figure}
  \begin{center}
    \plotone{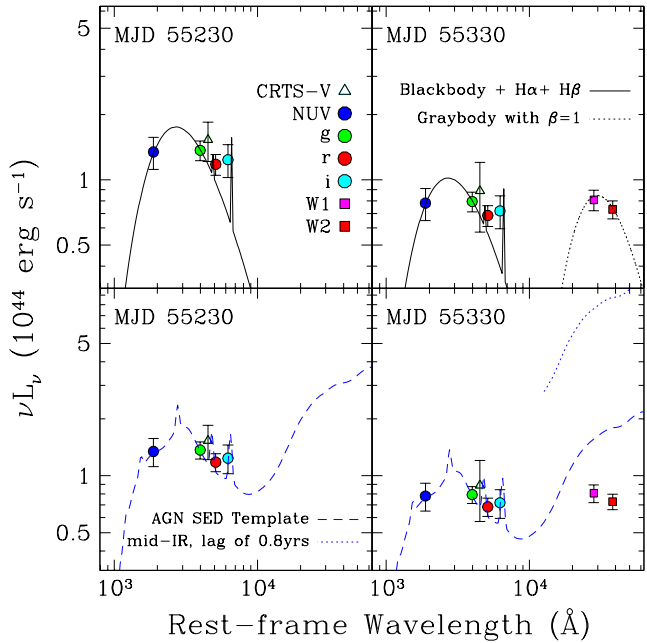}
    \caption{SED of the transient in \sname\ at MJD 55230 (left
      panels) and 55330 (right panels). The top panels show the
      best-fit blackbody and emission lines model (solid black lines)
      to the optical and UV bands, as discussed in the text. The
      dashed gray line shows the best-fit modified blackbody (or
      graybody) with $\beta=1$ to the WISE W1 and W2 bands. The bottom
      panels show the best-fit AGN SED model to the data (dashed blue
      line) for each epoch. In the bottom right panel, the dotted blue
      line shows the expected mid-IR emission if the transient had
      occurred within the accretion disk and the light travel time to
      the inner edge of the dust torus was 0.8~yrs, the upper limit we
      estimate for it.}
    \label{fg:SED}
  \end{center}
\end{figure}

Figure \ref{fg:SED} shows the continuum-subtracted specific luminosity
of the transient in the UV and optical bands for two epochs. The left
panels show MJD 55230, while the right panels show MJD 55330, one
hundred days later. MJD 55230, was chosen to be close to the first
epoch of Pan-STARRS observations and to the second epoch of the NUV
observations, thereby minimizing interpolation. To predict the
observed flux at each date for each band, we assume the same rate of
decay observed for the $V_{\rm CSS}$-band lightcurve after the
subtraction of the AGN emission. The apparent inconsistency observed
between the CRTS and the Pan-STARRS bands is most likely due to the
shape of the SED since the CRTS observations are unfiltered. The
second date shown in Figure \ref{fg:SED}, MJD 55330, corresponds to
the first epoch of the WISE observations. Note that the NUV and
Pan-STARRS bands shown here are estimated in the same manner as for
the MJD 55230 date. As there is no additional data, these estimates
are strongly dominated by the assumption of the decay rate being
consistent with that of the $V_{\rm CSS}$-band.

In the first date for which the SED is shown, MJD 55230, the NUV, $g$
and $r$ bands are qualitatively consistent with a blackbody whose peak
lies between the NUV and $g$ bands. The $i$-band specific luminosity
is much higher than would be expected for such a blackbody, but could
be driven by the H$\alpha$ emission line. If we assume the underlying
SED is composed by a blackbody continuum with strong H$\alpha$ and
H$\beta$ emission lines, we can fit all four data points perfectly, as
the model has exactly four parameters. For the emission lines, we
assume a Gaussian shape with the same width measured in the SDSS
spectrum (i.e., $\sigma=886.5~\rm km~\rm s^{-1}$). The best-fit model
is shown in the top left panel of Figure \ref{fg:SED}. For the later
date, MJD 55330, shown in the top right panel, we simply scale the fit
to the MJD 55230 data by the respective NUV specific luminosities. The
best-fit blackbody has a temperature of $13,600\pm 500~\rm K$ and the
H$\alpha$ and H$\beta$ emission lines have rest-frame equivalent
widths (EW) of $960\pm 340~\rm \AA$ and $240\pm 90~\rm \AA$,
respectively. Note that since these are broadbands, the EW is very
insensitive to the width assumed for the emission lines. For example,
if we assume a FWHM of $5000~\rm km~\rm s^{-1}$, the EWs change by
less than 1\%. Note that all the uncertainties on the best-fit
parameters are determined from a Monte Carlo process that considers
the uncertainty on the observed fluxes as well as the uncertainty in
the continuum subtraction. Note also that the emission of the WISE
bands is inconsistent with this blackbody model, requiring a different
component to describe them. We model the WISE bands by a modified
blackbody (or graybody) with $\beta=1$, and find an implied
temperature of 970~K. The fit is shown in Figure \ref{fg:SED}, but we
note that the temperature is highly unconstrained due to the fact that
we fit two parameters of the model (amplitude and temperature) to two
data points.

Alternatively, we could assume that the underlying SED of the source
is still that of an AGN. While the analysis of \citet{graham17}
determined that the source of the transient cannot be regular AGN
variability, it could be the case that the transient event maintains
the AGN SED (we discuss this further in
\S~\ref{sec:transient_source}). The bottom panels of Figure
\ref{fg:SED} show the best-fit obtained in this case for the two
epochs using the AGN SED template and algorithm of \citet{assef10},
without a host galaxy component. The fit to the UV and optical bands
has a $\chi^2_{\nu}=0.07$ on both dates. The best-fit model requires a
reddening of $E(B-V)=0.16\pm 0.01$, which is considerably larger than
that found for the pre-transient SED of \sname. While this could be
taken as an indication that the SED of the transient deviates from
that of the AGN in \sname, it is also possible that the nuclear
obscuration could have changed between the two epochs.

The best-fit SED at MJD 55330 significantly overpredicts the observed
W1 and W2 fluxes, which is difficult to reconcile with the source of
the transient being within the accretion disk. As the emission of the
dust torus would simply be reprocessed UV/optical emission from the
accretion disk and the transient in this scenario, the excess dust
emission would have to be commensurate with the transient emission
with a time lag given by the light travel time to the inner edge of
the torus (which is $<$0.8~yrs). We estimate that 0.8~yrs before MJD
55330, the $V_{\rm CSS}$ emission of the transient would be 4.9 times
brighter. For comparison, the bottom-right panel of Figure
\ref{fg:SED} shows the expected dust emission reprocessing 4.9 times
brighter UV/optical emission. If the transient had been located within
the accretion disk, we would have expected the W1 and W2 fluxes to
have been within the two curves shown. That the observed dust emission
is significantly fainter instead implies that this physical scenario
is not viable.

\subsection{Energy of the Transient Event}\label{ssec:energy}

As discussed in the previous section, the $V_{\rm CSS}$-band and the
WISE W1 and W2 bands show different behaviors with time, with the WISE
bands fading less quickly than the optical light. This suggests that
the WISE bands are tracing a different physical component of the
transient. Since the mid-IR emission is likely related to a dusty
component that reprocess the optical emission, we estimate the energy
of the transient event by using the $V_{\rm CSS}$-band lightcurve
alone. Because the lightcurve is somewhat noisy and sparse, we group
the $V_{\rm CSS}$-band data in bins of 10 days, and we linearly
interpolate the binned lightcurve in magnitude space to determine the
luminosity as a function of time.

We first assume the best-fit blackbody discussed in the previous
section. Note that while that model also considered strong H$\alpha$
and H$\beta$ emission lines, we do not count their contribution to the
luminosity for the energy estimate. In order to estimate the total
radiated energy, we scale the luminosity of the best-fit model to
match the flux at a given date, but we do not allow the temperature to
evolve with time. Since the exact filter curve of the $V_{\rm
  CSS}$-band is not known, we rely on the empirical calibrations of
\citet{drake13} between $B$, $V$, $R$ and $I$, and the $V_{\rm
  CSS}$-band\footnote{A typographical error is present in eqn. (3) of
  \citet{drake13}. The correct equation is $V = V_{\rm CSS} + 1.07
  \times (R-I)^2 + 0.04$ (Drake, A.~J., private comm.).}  to estimate
the black body luminosity at a given time. In practice, we use the
{\tt{PySynphot
    v0.9.7}}\footnote{\url{https://pysynphot.readthedocs.io/en/latest/}}
package to compute the observed $B$, $V$, $R$ and $I$ magnitudes for
the SED model described, and use these calibrations to estimate its
$V_{\rm CSS}$-band magnitude, and then scale its luminosity to fit the
observations. Note that for each set of $B$$V$$R$$I$ magnitudes, the
calibrations of \citet{drake13} naturally yield three estimates of the
$V_{\rm CSS}$-band, and hence three estimates of the
luminosity. Considering this, as well as the uncertainities in the
best-fit parameters of the SED model and the uncertainties in the DRW
subtraction of the underlying continuum, we estimate the total
radiated energy of the transient event to be $E=1.6\pm 0.3\times
10^{52}~\rm erg$.

As discussed in the previous section, the SED of the transient is
somewhat consistent with that of an AGN. If instead of using the
blackbody SED model, we obtain the luminosity by integrating over the
best-fit AGN SED model and repeat the above process, we estimate a
total radiated energy of $E=1.0\pm 0.2\times 10^{53}~\rm
erg$. However, in order for this estimate to be accurate, we need to
remove the contribution of the dust torus to the luminosity. Assuming
it accounts for about half of the integrated luminosity of the SED
template (after removing the reddening), the energy estimate would be
$\sim 5\times 10^{52}~\rm erg$.

\section{Source of the Transient Event}\label{sec:source_w0948}

\citet{graham17} discussed a number of sources similar to
\sname\ found in the CRTS survey. They explored in detail whether
these transients could be microlensing events caused by stars in the
Milky Way, super-luminous supernovae, TDEs, or black hole binary
mergers. \citet{graham17} finds that the majority of the CRTS
transients are incompatible with microlensing events and, as discussed
earlier, microlensing of the central engine in \sname\ would not be
observable in the WISE bands, given that the physical scale of the
dust torus is much bigger than that of the accretion disk. Hence, we
do not discuss this scenario any further. Below, we first compare the
lightcurve of the transient in \sname\ with those of other energetic
transient events, and then we discuss its nature in the context of
other cases outlined by \citet{graham17}. We do not discuss stellar
mass binary black hole mergers within the accretion disk of the AGN,
as although intriguing, no testable predictions have been
made. Additionally, we discuss a mechanism recently proposed by
\citet{moriya17} to explain similar transients in AGN.

\subsection{Comparison with Other Energetic Transient Events}\label{ssec:lc_comp}

\begin{figure}
  \begin{center}
    \plotone{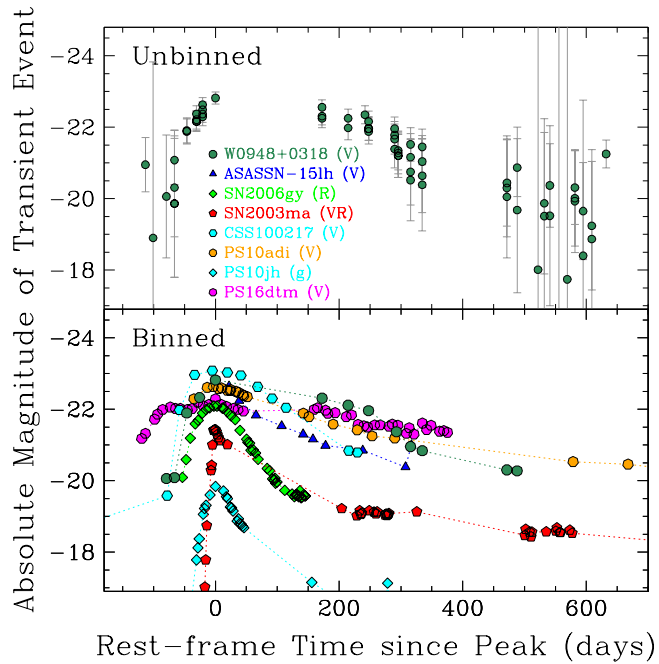}
    \caption{({\it{Top}}) The green circles shows the CRTS $V_{\rm
        CSS}$-band lightcurve of the transient in
      \sname. ({\it{Bottom}}) The green circles show the same as in
      the top panel but binned in groups of 10 days. The Figure also
      shows a number of different luminous transients in the
      literature for comparison with the transient in \sname.}
    \label{fg:comparison}
  \end{center}
\end{figure}

Figure \ref{fg:comparison} compares the lightcurve of the transient
event in \sname\ with a number of very luminous transients in the
literature. The Figure shows the lightcurve for the luminous transient
event CSS100217 discovered by \citet{drake11} using CRTS, with a total
radiated energy of $1.3\times 10^{52}~\rm erg$. The peak absolute
$V_{\rm CSS}$-band magnitude of this event is similar to that of the
one in \sname, although it has a faster decay rate. Interestingly, the
transient occurred in a narrow-line Seyfert 1 galaxy, similar to the
host of \sname\ (see \S\ref{ssec:sdss_spec}). \citet{drake11} discuss
a number of scenarios for this transient and find that the most likely
is that CSS100217 corresponds to a type IIn supernova that occurred
within the narrow-line region of the AGN. We also show the luminous
transient event PS16dtm (or SN 2016ezh), also hosted in an AGN
(specifically a narrow-line Seyfert 1). \citet{blanchard17} shows that
the lightcurve of this object is very flat in a number of different
optical bands, and determines that the transient is best modeled by a
TDE. The $g$-band lightcurve of the significantly the less luminous
TDE, PS10jh \citep{gezari12}, is also shown. Unlike the previous
objects, PS10jh has a significantly faster decay time, and it is not
hosted in an AGN.

Figure \ref{fg:comparison} also shows the $V$-band lightcurve of
ASASSN-15lh. Discovered by \citet{dong16}, this is the most luminous
supernova ever detected, with a total radiated energy of
$1.7-1.9\times 10^{52}~\rm erg$ within its first 550 days
\citep{godoy17}. \citet{dong16} determined this transient to be a
hydrogen poor super luminous supernova \citep[although see][who have
  suggested this transient to be a TDE from a Kerr black
  hole]{leloudas16}. The decay rate of this event is quite similar to
that of the transient in \sname\ and likely has a similar peak
magnitude. Additionally, the Figure shows the $R$-band lightcurve of
the luminous hydrogen-rich supernova SN2006gy from \citet{smith07},
for which they determined a total radiated energy of $1.2\times
10^{51}~\rm erg$. The decay rate of this transient is much faster than
that of the transient in \sname, suggesting a different physical
origin. A different luminous type IIn supernova, SN2003ma
\citep{rest11}, is also shown and while it has a significantly lower
peak magnitude than either SN2006gy or \sname, it has a slower decay
rate, similar to that observed for the transient in
\sname. \citet{rest11} estimate the total radiated energy of SN2003ma
to be $3.6\pm 1\times 10^{51}~\rm {erg}$.

Recently, \citet{kankare17} studied the very luminous transient
PS10adi, whose host galaxy is an AGN. The total energy radiated by the
transient is estimated to be $\sim2.3\times 10^{52}~\rm erg$, somewhat
above, but comparable to, the energy we estimated for the transient in
\sname. \citet{kankare17} concluded that this transient was consistent
with a TDE or a superluminous type II SN within the narrow-line region
of the AGN. The lightcurve of PS10adi is also shown in Figure
\ref{fg:comparison}, and it can be seen to be very similar to that of
the transient in \sname, suggesting a common origin. In fact,
\citet{kankare17} identified \sname\ as a possible analog of
PS10adi. In the next sections we'll discuss the possible sources for
the transient in \sname, and find that it is unlikely for the
transient to be a TDE or a type II SN within the accretion disk of the
system, but instead favor a location within the dust torus of the AGN.

\subsection{Super-Luminous Supernova}\label{sec:transient_source}

Given the high peak luminosity ($M_V\sim -23$) and long-lasting
lightcurve, if the transient in \sname\ was due to a supernova, it
would have to be a super-luminous supernova
\citep[SLSN;][]{galyam12}. As discussed earlier, the super-luminous
SNe ASASSN-15lh (hydrogen poor), SN2003ma and CSS100217 (hydrogen
rich) all show comparably slow (albeit slightly quicker) decay rates
to the transient in \sname. If the source of the transient in
\sname\ were a SLSN, we would expect that the underlying SED would be
generally consistent with a blackbody. As discussed in
\S\ref{ssec:transient_sed}, the SED of the transient in \sname\ is
indeed consistent with a blackbody, as long as we allow for
significant H$\alpha$ and H$\beta$ emission, which are also expected
for SNe in general. Assuming this SED, in \S\ref{ssec:energy} we found
a total radiated energy of $E=1.6\pm 0.3\times 10^{52}~\rm erg$. This
is consistent with the total radiated energy of $1.3\times10^{52}~\rm
erg$ estimated for CSS100217 by \citet{drake11}, but significantly
larger than the $3.6\times 10^{51}~\rm erg$ estimated for SN2003ma by
\citet{rest11}.

The total radiated energy for ASASSN-15lh was estimated by
\citet{godoy17} to be $1.7-1.9\times 10^{52}~\rm erg$ within the first
550~days after first detection, which is consistent with the value
estimated for the transient in \sname. \citet{dong16} also found the
best-fit blackbody to ASASSN-15lh had temperatures ranging from
21,000~K at 15 days after the peak, to 13,000~K at 50 days after the
peak. \citet{godoy17} found that after $\sim$200 days the event
started rebrightening and the temperatures started increasing again
with time, from a minimum of 11,000~K to a maximum of 18,000~K. For
the transient in \sname, we found a best-fit black body temperature of
$13,600\pm 500~K$ at $\sim$250~days after the peak, consistent with
the range found in ASASSN-15lh. We note that our energy estimate
assumes a single temperature throughout the duration of the
transient. Significantly lower temperatures would not have a major
effect over the estimated luminosities at each date, as the bulk
emission of such a blackbody would be within the wavelength range
covered by the CRTS $V_{\rm CSS}$-band. On the other hand,
significantly higher temperatures would strongly increase the
estimated luminosity at each date, since the $V_{\rm CSS}$-band would
only cover the Rayleigh-Jeans tail of the continuum emission. Hence,
if the transient in \sname\ has the same physical origin as
ASASSN-15lh, then our estimate for the total energy radiated would be
a conservative lower bound.

As mentioned above, we find that in order to explain the SED we need
H$\alpha$ and H$\beta$ emission lines with equivalent widths of
$960\pm 340~\rm \AA$ and $240\pm 90~\rm \AA$ respectively, which are
compatible with what has been observed in luminous SNe. For example,
\citet{jencson17} observed that in the luminous type II SN SN2010jl
the H$\alpha$ EW varied between 170~\AA\ and 3500~\AA\ throughout the
lifetime of the transient, but was approximately constant at
$\sim$2000~\AA\ at late times ($\gtrsim$400~days after first
detection), a similar timescale at which we can model the SED of
\sname. \citet{jencson17} also observed that the H$\beta$ EW varied
between 35~\AA\ and 300~\AA.

Hence, we conclude that the transient in \sname\ is highly compatible
with a SLSN. For hydrogen poor (or type I) SLSN, the transient is
thought to be related to the formation of a magnetar during a SN
explosion, where the increased luminosity arises from the coupling
between the magnetar spindown energy and the SN ejecta
\citep{kasen10,woosley10}. For type II (hydrogen rich) SLSN the
leading mechanism for its radiated energy, instead, is thought to come
from the interaction of a SN explosion ejecta with a dense
circumstellar medium \citep[CSM; e.g., see][]{smith07,smith14}. Such a
dense CSM could naturally be found within or in the vicinity of an
AGN. In fact, \citet{sukhbold16} argue that total radiated energies in
excess of $\sim 3\times 10^{51}~\rm erg$ are not attainable by type II
SLSN if the CSM only comes from material ejected by the parent star
prior to the explosion. Furthermore, the large amount of hot dust
emission observed during the transient in \sname\ suggests the
transient occurred within a very dusty medium, which could be
consistent with the dust torus of the AGN. As discussed in
\S\ref{ssec:transient_sed}, it is unlikely that the SLSN would be
within the accretion disk, as the W1 and W2 magnitudes do not seem to
lag the optical magnitudes with a long enough timescale. However, the
fact that we see little to no evolution in the W1--W2 color throughout
the transient, and at later times, implies the dust is not changing
temperature, and could be consistent with the transient occurring
within the torus, as the accretion disk emission would maintain the
high temperature regardless of the transient luminosity. The high gas
density in the torus could provide an ideal CSM to generate high
luminosities.

We note that \citet{drake11} came to a similar conclusion for
CSS100217, i.e., a SN within the immediacies of the AGN. The
similarities in energy and lightcurve suggest they may have a similar
origin, but unfortunately there were no time-resolved mid-IR
observations of the CSS100217 transient to compare with the optical
lightcurve.

\subsection{Accretion Disk Flares, Tidal Disruption Events and Accretion Disk Outflows}

As discussed in \S\ref{ssec:transient_sed}, the WISE lightcurves are
inconsistent with the transient occurring in the accretion disk of the
AGN, which in turn implies the transient in \sname\ is unlikely
related to a TDE. Furthermore, the optical SED itself also suggests
that a TDE is an unlikely source for the transient. While the majority
of TDE events identified in the literature have significantly lower
peak luminosities than the transient in
\sname\ \citep[e.g.,][]{gezari12,holoien14,holoien16a,holoien16b,brown18},
these have all been in galaxies that do not host nuclear activity,
implying there is significantly less material for the ejecta to
interact with and hence increase the total radiated energy
output. However, recently \citet{blanchard17} identified the TDE
candidate PS16dtm, whose host galaxy has an AGN similar to that of
\sname. This TDE is significantly more luminous than all previous TDE
candidates. While the peak magnitude of this event is $\sim 1~\rm mag$
fainter than the transient in \sname, the event is extremely
energetic, owing in great part to its almost constant lightcurve
$\sim$100 days after peak emission. Unfortunately there is a gap of
$\sim$150 days after the peak in the CRTS observations of \sname, so
we cannot fully assess the similarity of the lightcurves in this
period. Yet, the lightcurve of the transient in \sname\ seems to be
highly inconsistent with such behavior in Figure
\ref{fg:comparison}. We note that \citet{kankare17} find that PS16dtm
is a possible analog to PS10adi, just as the they find for \sname. We
also note that recently \citet{mattila18} observed a dust obscured TDE
in the nucleus of the nearby galaxy Arp 299, with an estimated total
radiated energy $>1.5\times 10^{52}~\rm erg$, and likely in the range
$(1.9-6.5)\times 10^{52}~\rm erg$ when considering dust re-radiation
at longer IR wavelengths not covered by their observations. This is
somewhat larger than, yet comparable to, the total radiated energy
estimated for the transient in \sname. However, the
{\it{Spitzer}}/IRAC observations of the event in the [4.5] band show
that the brightness of the transient falls approximately 0.9~mag in
the 2000 days after the peak (or $4.6\times 10^{-4}~\rm mag~\rm day
^{-1}$), an order of magnitude more slowly than found for the W2 band
lightcurve of the transient in \sname\ in \S\ref{ssec:transient_sed},
suggesting they might be powered by different mechanisms.

Recently \citet{moriya17} proposed a mechanism to explain transients
in AGN such as CSS100217 and PS16dtm, through the interaction of mass
ejections from the accretion disk instead of unrelated explosive
phenomena such as SNe or TDEs. Specifically, they ascribe the observed
transients to the interaction between mass outflows from the accretion
disk and the broad-line region (BLR). They determined that a mass
outflow of $1~M_{\odot}$ can produce a total radiated energy of
$10^{52}~\rm erg$ over timescales of a few hundred days. Unfortunately
no specific predictions are made about the lightcurve, and hence it is
difficult to test how well the transient in \sname\ can be described
by this model. In the specific case of CSS100217, \citet{moriya17}
describes that these transients may be reoccurring in timescales of
years to decades. Given the similarities between the two cases,
continuous monitoring of this source would be necessary to confirm
this scenario. However, as the radius of the BLR, $R_{\rm BLR}$, in
\sname\ would be at $\sim$25~light days from the accretion disk
according to the AGN's 5100\AA\ luminosity and the $R_{\rm
  BLR}-\rm L_{5100}$ relation of \citet{bentz09}, the low mid-IR
emission compared to the optical emission observed during the
transient (see \S\ref{ssec:transient_sed}) may be hard to reconcile
with this scenario.

\section{Obscured Analogs to \sname}\label{sec:analogs}

As discussed in \S\ref{ssec:lightcurve} and \S\ref{ssec:energy}, the
mid-IR emission is likely tracing a different physical component than
the optical emission. This component is likely a dusty structure
reprocessing the optical light emission of the transient. This
structure could be either the dust torus if the source of the event
was the accretion disk itself or a star close to it, or circumstellar
dust in the case the source was a supernova in the host galaxy. In
\S\ref{ssec:var_agn_lightcurves} we showed that none of the other
optical lightcurves for the objects in our sample are qualitatively
similar to that of \sname. However, because of the nearby dust
reprocessing the optical emission, it is possible that analogs to
\sname\ exist within our sample but are simply dust obscured.

We search for transients with similar WISE lightcurves to
\sname\ among our sample. In order to identify the candidates, we use
the following criteria, based on the lightcurve of \sname:
\begin{itemize}
  \item[a.] The first two years of the NEOWISE-R data should not be
    highly variable. Specifically, we require that a constant flux fit
    to the NEOWISE-R data yields a reduced $\chi^2$ of $<3$ within its
    first four epochs. We do not extend this to the later data as AGN
    are intrinsically variable in the mid-IR in long timescales
    \citep[e.g.][]{kozlowski16}.

  \item[b.] The lightcurve must show excess flux during the AllWISE
    observations as compared to the NEOWISE-R ones. In practice, we
    require that the median magnitude of the last AllWISE epoch is
    brighter than the median magnitude of the first NEOWISE-R epoch.
\end{itemize}

\begin{figure}
  \begin{center}
    \plotone{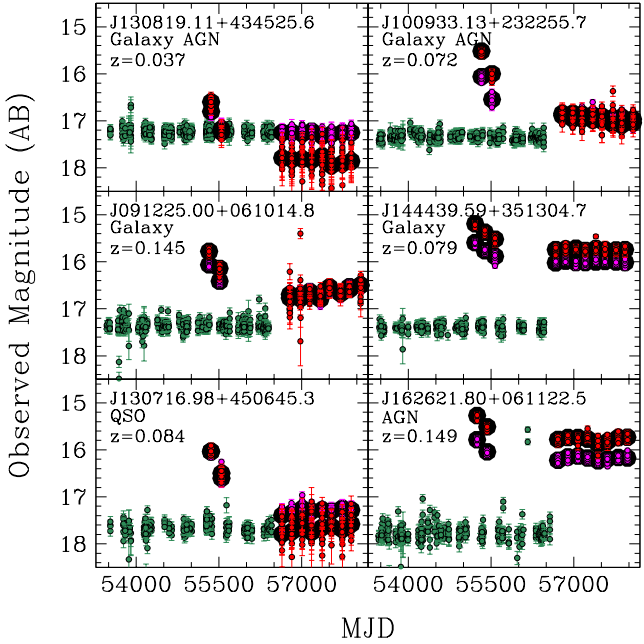}
    \caption{Lightcurves of the six brightest analogs to
      \sname\ identified in \S\ref{sec:analogs}. The sources are
      arranged in descending brightness order. The symbols have the
      same meaning as in Fig. \ref{fg:lcs_high_var}.}
    \label{fg:analogs1}
  \end{center}
\end{figure}

\begin{figure}
  \begin{center}
    \plotone{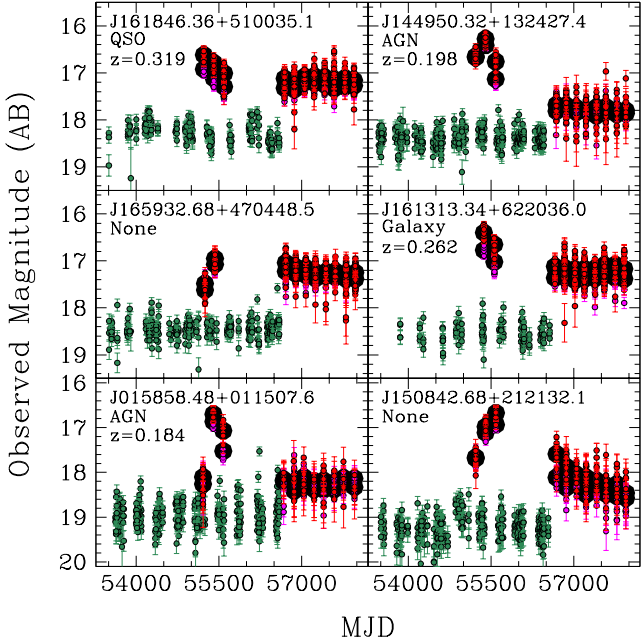}
    \caption{Lightcurves of the six faintest analogs to
      \sname\ identified in \S\ref{sec:analogs}. The sources are
      arranged in descending brightness order. The symbols have the
      same meaning as in Fig. \ref{fg:lcs_high_var}.}
    \label{fg:analogs2}
  \end{center}
\end{figure}

These criteria results in 12 candidates, in addition to \sname, out of
the sample of 45 highly variable mid-IR AGN identified by
\citet{assef18}. These 12 candidates are shown in Figures
\ref{fg:analogs1} and \ref{fg:analogs2}. Most of the objects show
mid-IR lightcurves that are highly compatible with that of \sname,
although none of them show any significant optical variability in
CRTS. If these transients have the same origin as in \sname, then it
is further unlikely that the accretion disk is the source of the
transient, as its optical emission would have been detected by CRTS in
at least the two sources that are spectrally classified as QSOs by
SDSS, namely WISE J130716.98+450645.3 ($z=0.084$) and WISE
J161846.37+510035.2 ($z=0.319$). In those objects we have a direct
line of sight towards the accretion disk that is not obscured by the
dust torus, unless they have significantly changed their optical
obscuration since when the spectroscopic observations were
obtained. In order to test this possibility, spectroscopy of the
reminder of the sample is being gathered and those will be later
reported by \citet{jun17}.

If we apply the same selection criteria to the sample of
\citet{graham17} discussed in \S\ref{ssec:var_agn_lightcurves} and we
further request that the AllWISE observations have {\tt{var\_flg}}$=$9
in both W1 and W2, we find that \sname\ is the only source
selected. If we relax the latter requirement to only require
{\tt{var\_flg}}$\ge$7 in both bands, i.e., that source is a real
variable in W1 and W2 during AllWISE, we find one additional source,
CRTS0234+0107, would be selected as an analog to \sname. The optical
and mid-IR lightcurve of this source, along with those of the rest of
the \citet{graham17} sample, are shown in the Appendix
\ref{app:graham}.

\begin{figure}
  \begin{center}
    \plotone{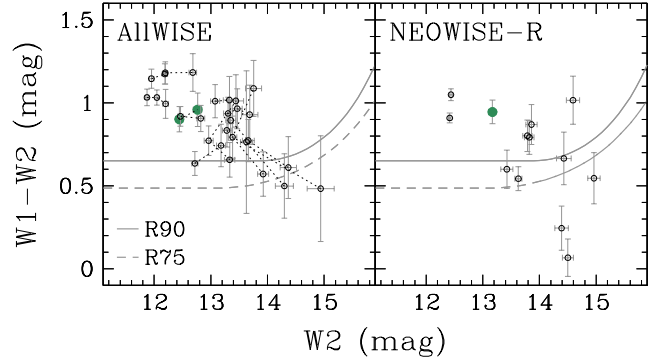}
    \caption{W1--W2 color against W2 magnitude for \sname\ (large
      green circles) and its possible analogs (small open
      circles). The left panel shows the quantities during each
      AllWISE epoch, where different epochs are connected by the
      dotted lines. The right panel shows the average values of these
      quantities during the NEOWISE-R epochs for each object. In each
      panel, the solid (dashed) gray line shows the R90 (R75) AGN
      selection criterion of \citet{assef18}.}
    \label{fg:agn_criteria_candidates}
  \end{center}
\end{figure}

In \S\ref{ssec:lightcurve} we discussed that the W1--W2 color of
\sname\ did not change between the AllWISE and NEOWISE-R observations,
and was consistent with the AGN classification of \citet{assef18} at
all times. In Figure \ref{fg:agn_criteria_candidates} we show the W2
magnitude and W1--W2 color of each AllWISE epoch, and the mean of the
NEOWISE-R epochs. We also show the R90 AGN selection criteria of
\citet{assef18} as well as their less restrictive (but lower
reliability) R75 selection. All objects redder than the R90 or R75
boundaries are considered to be AGN. During the AllWISE epochs, all
objects are consistent with the R90 classification at least within the
errorbars, although some show a clear transition into bluer, and hence
less AGN dominated, mid-IR SEDs. For the NEOWISE-R observations
however, there are four objects significantly bluer than the R90
criteria. Two of these are consistent, at least within the errorbars,
with the R75 criteria, and hence might still be bona-fide AGN. The
other two would certainly not be classifiable as AGN: WISE
J130716.98+450645.3 (W1307+4506) and WISE J130819.12+434525.6
(W1308+4345). Notably, both have spectroscopic observations from SDSS
and both are classified as AGN. Specifically, W1307+4506 is classified
as a QSO at $z=0.084$ with clear broad H$\alpha$ and H$\beta$
profiles, while W1308+4345 has narrow-line ratios that imply it is a
type 2 AGN. The change in WISE colors imply that while during the
AllWISE observations the W1 and W2 bands were dominated by the hot
dust emission from the AGN torus, in the NEOWISE-R observations they
are dominated by the stellar emission from the host galaxy. Hence, the
AGN in these objects were caught by the AllWISE observations possibly
migrating into a lower accretion state and could be classified as
changing look AGN. Indeed, \citet{assef18} classified WISEA
J142846.71+172353, one of the other high variability WISE AGN
candidates (although not an analog to \sname), as a changing look
quasar. Follow-up spectroscopic observations to be presented by
\citet{jun17} will be able to better constrain the nature of these two
targets.

W1308+4345 is one of three of the possible analogs to \sname\ objects
that have recently being identified as possible TDE candidates by
\citet{wang18}. The other two objects are WISEA J091225.00+061014.8
(W0912+0610) and WISEA J100933.13+232255.7 (W1009+2322). We note that
W1009+2322 and W1308+4345 are classified as Galaxy AGN by SDSS,
although \citet{wang18} instead classifies them as star-forming
galaxies.  W0912+0610 is identified by SDSS as a galaxy without an AGN
component. If these objects are indeed TDEs, then they are unlikely
analogs to \sname.

For the remaining targets, there are a number that are classified as
type 1 AGN or QSOs that do not show any optical variability associated
with the mid-IR transient. In the previous section, we concluded that
the most likely explanation for the transient in \sname\ is a SLSN,
and we speculated that due to the near-constant mid-IR color
throughout and after the transient event, that the SLSN could have
happened within the torus of the AGN, and that we happen to have an
unobscured line of sight towards the event. If we assume that the
majority of these analogs share a similar origin to the transient in
\sname, the lack of optical transients is consistent with locations
within the tori of their AGN. In other words, in the majority of these
events, the transient would remain obscured and only observable in the
mid-IR, with the transient in \sname\ being the exception on which an
unobscured line of sight was available.

\subsection{Rates}

Determining an event rate is difficult in this case as we have very
little data to characterize the transient, and there are a large
number of selection effects that went into the identification of
\sname. The parent sample consisted on 4,544,197 AGN candidates from
the R90 catalog of \citet{assef18}, which consider all W1 and W2
selected AGN in 30,093~deg$^{2}$ of the AllWISE data release. This
area corresponds to the entire extragalactic sky, avoiding the
Galactic Center, the Galactic Plane, nearby galaxies and known
star-formation regions in our Galaxy \cite[see][for
  details]{assef18}. \citet{assef18} required all their candidates to
be point sources in the WISE images, which puts a significant bias
against the most nearby objects.

Furthermore, the selection of variable objects required them to be
identified with the highest variability level in the AllWISE source
catalog. This criteria is based upon the probability of the source
being variable given its fluxes and uncertainties in the single frame
observations, and generates two biases. The first is that the number
of single frame observations is a steep function of the ecliptic
latitude due to the survey scan pattern of WISE. This means that
sources at higher ecliptic latitudes have better statistics from which
to compute this variability index. The second bias is against fainter
objects, as their single exposure photometric uncertainties are
larger, requiring a higher amplitude of the variability to be
cataloged with the highest variability level. In fact, \sname\ has
coadded magnitudes in AllWISE of W1$=$13.537 and W2$=$12.603, while
the median magnitudes of the R90 catalog of \citet{assef18} are
respectively 16.58 and 15.42. Only 0.4\% (18,823) of the sources are
brighter than \sname\ in W1, and only 0.6\% (28,242) are brighter in
W2. To have a good variability detection, the sources have to be well
detected in the single frame exposures, which requires
W1$<$14.25~mag\footnote{\url{http://wise2.ipac.caltech.edu/docs/release/allwise/expsup/sec2\_3c.html}}. For
fainter W1 magnitudes the variability detection is severely
degraded. In fact, of all the analogs of \sname\ identified in the
previous section, only WISEA J165932.68+470448.5 has a peak magnitude
fainter than this limit, with a median W1$=$14.39 in the brightest
AllWISE epoch. Further inspection, however, shows that the peak single
frame magnitude measured for this objects is W1$=$11.56, with the
second brightest magnitude being W1$=$14.26. As the AllWISE database
does not show any other targets within 3\arcsec\ of WISEA
J165932.68+470448.5, the assignment of the highest variability flag
could be spurious. Hence, we do not consider this object further in
our analysis. 

In the NEOWISE-R epochs, the mean magnitude of \sname\ in W1 is 14.12,
or $\sim$0.77~mag fainter than the median W1 magnitude in the first,
brightest AllWISE epoch. Hence, we can assume that we should have been
able to detect a transient like the one in \sname\ in any host with an
apparent brightness down to 0.77~mag fainter than the W1$=$14.25~mag
limit discussed above, i.e., with W1$\lesssim$15.02~mag. In the R90
catalog of AGN candidates of \citet{assef18} there are 338,861 such
objects, and hence we can use this number to estimate the rate of
these transient events. This is a conservative assumption, however, as
it is unlikely that the object would have been identified as a highly
variable AGN by WISE so close to the detection limit, in particular
without having been detected in more than one epoch.

Considering all of these sources are at low redshifts, and hence to
zero-th order we can ignore the cosmological time dilation, we can get
a lower bound on the estimate of the occurrence of these
transients. The duration of the optical transient in \sname\ is
$\sim$800~days, counting from the peak emission to the point where the
CRTS lightcurve becomes dominated by the host emission. Since the
AllWISE survey has a span of approximately 180~days between epochs, we
probe a total period of $\sim$1,000~days, meaning we could have
detected from transients that started $\sim$800 days before the first
epoch up to transients that started on the second AllWISE epoch. This
assumption may oversimplify the estimate, as it implies we could have
detected the transient at any stage of its evolution. Disregarding
this issue, we can estimate that the rate per AGN of these transients
would be approximately $1/338,861/(1000~\rm days) =
1.1^{+2.5}_{-0.9}\times 10^{-6}~\rm yr^{-1}$, or about once in every
$900,000^{+4,400,000}_{-640,000}$ years in a given AGN if \sname\ is a
unique object within our sample (i.e., disregarding all possible
analogs presented in the previous section). The error bars consider
the uncertainty derived from a Poisson distribution
\citep{gehrels86}. However, because of the simplifications made, this
should be considered as a lower bound estimate. This becomes further
the case when we consider that for a fixed energy of the transient,
the effects will be significantly less obvious in more luminous AGN,
and hence we might not see them. We can then only conclude that the
detection implies that these transients happen with a rate greater
than $>2\times 10^{-7}~\rm yr^{-1}$ per AGN, or more than once every
$\sim$5 million years. Considering that the lifetime of an AGN is
$\gtrsim$10 million years \citep{martini01,marconi04}, the above rates
suggest that these transient events would occur $\gtrsim$2 times
during the phase on which the central supermassive black hole is
actively accreting.

Similarly, we can consider instead the event in isolation, and
consider we could have detected the transient if it was up to a
redshift on which the apparent peak magnitude at W1 would have been
14.25~mag. Neglecting K-corrections, we find that the redshift at
which this apparent magnitude would have been observed is
$z=0.290$. Considering that the R90 catalog of \citet{assef18} covers
30,093~deg$^2$ of the sky, we find that the event was detected within
a comoving volume of 5~Gpc$^3$. This implies a rate of
$0.07^{+0.17}_{-0.06}~\rm Gpc^{-3}~\rm yr^{-1}$. As per the discussion
above, this number is likely a lower limit, and hence we conclude that
the rate is $>0.01~\rm Gpc^{-3}~\rm yr^{-1}$. If we compare with the
rate of type II SLSN of $151^{+151}_{-82}~\rm events~Gpc^{-3}~\rm
yr^{-1} h^3_{71}$ found by \citet{quimby13}, we find that the
transients in \sname\ are less frequent by a factor of
$<151,000$. \citet{quimby13} finds a rate of type I SLSN of
$68^{+94}_{-44}~\rm events~Gpc^{-3}~\rm yr^{-1} h^3_{71}$, implying
the rate of transients like the one in \sname\ are less frequent by a
factor of $<6,800$. This implies that we cannot, in principle,
disregard the possibility that the transient in \sname\ is a SLSN
unassociated with the AGN. If this had been the case then we would
have expected that the rates would differ by a factor of $\sim$1,000,
as roughly 1 in every 1,000 galaxies hosts an active nucleus, which is
consistent with the estimates above. However, the transients studied
by \citet{quimby13} have lower radiated energy than the transient in
\sname, and if there is a steep dependence of the rates on energy, as
suggested by \citet{godoy17}, the rates could be hard to reconcile.

In the previous section we presented 12 additional objects whose WISE
lightcurves suggest they could be analogs to \sname, although two of
them showed W1--W2 colors during the NEOWISE-R observations no longer
consistent with an AGN classification, unlike \sname, and one may have
a spurious variability classification. If we assume that the remaining
9 transient candidates are obscured analogs to the transient in
\sname, their rate per AGN would instead be $>7.5\times 10^{-6}~\rm
yr^{-1}$, or greater than once every 135,000~years, about 30 times
more frequent. Additionally, we can estimate the event rate density by
considering, as done above, the maximum volume in which we could have
identified the targets. Excluding WISEA J150842.68+212132.1, for which
we do not have a redshift, we find a rate of $>1.2~\rm Gpc^{-3}~\rm
yr^{-1}$. This would imply these transients are a factor of $<125$
($<57$) times less frequent than type I (II) SLSN. These numbers are
somewhat inconsistent with the estimate of \citet{quimby13} if we
assume 1 in every 1,000 galaxies hosts an AGN, pointing to a relation
between AGN activity and these kind of transient events, and further
supporting the scenario discussed earlier of SLSN preferentially
occurring in the tori of AGN.

\section{Conclusions}

\citet{assef18} recently presented two catalogs of WISE-selected AGN
over 30,093~deg$^2$. From their catalog optimized for sample
reliability with $\sim$4.5 million AGN candidates, \citet{assef18}
noted that 687 were classified with the highest variability level in
the AllWISE database. \citet{assef18} found that for the 207 of those
targets located within the FIRST survey footprint, 162 (78\%) were
detected at radio energies, implying they are most likely
blazars. Here, we have presented a detailed study of the 45 highly
variable WISE AGN candidates that are undetected by
FIRST. \citet{assef18} discussed the spectroscopic classification of
34 of these sources, primarily from literature observations. They
found 31 of the 34 targets were consistent with an AGN
classification. In \S\ref{ssec:var_agn_spec_class} we presented
spectra for five additional sources, and found four of them to have
clear AGN signatures.

Although these 45 sources are highly variable in the WISE bands during
the AllWISE mission, we find that only seven show significant optical
variability in the CRTS survey, which covers the complete time-span of
the AllWISE observations (see \S\ref{ssec:var_agn_lightcurves} for
details). Two of them are carbon stars, and are contaminants to our
AGN sample. The other five have spectroscopic AGN
classifications. Four show low amplitude optical variability while the
fifth one, \sname, shows a very bright optical transient, coincident
in time with the AllWISE epochs of observation. \sname\ is
spectroscopically classified by SDSS as a QSO at $z=0.2073$. From the
SDSS optical spectrum, obtained significantly before the transient, we
find an SMBH mass of $M_{\rm BH} = 1.2\pm 0.4\times 10^{7}~M_{\odot}$
and that the AGN in \sname\ had an Eddington ratio at the time of
$0.25\pm 0.14$. From GALEX, SDSS and 2MASS broad-band photometry
obtained before the transient event, we find an SED that is best
modeled as a type 1 AGN with very little, and possibly no, reddening,
but with a significant host component.

In order to isolate the emission from the transient event from that of
the AGN itself in the CRTS $V_{\rm CSS}$ band, we modeled its
lightcurve before and after the transient event with a DRW model, and
subtracted the mean expected AGN emission during the transient
event. We find that the transient event is very asymmetric, with an
optical rise time of 0.019~mag~day$^{-1}$, $\sim$3 times faster than
the decay rate of 0.0059~mag~day$^{-1}$. We used the best-fit DRW
model to the $V_{\rm CSS}$ band to also isolate the transient event
emission in multi-wavelength broad-band observations from GALEX in the
NUV band and from PanSTARRS in the $g$, $r$ and $i$ bands. To isolate
the transient emission in the WISE bands, we subtracted from the
AllWISE bands the median of the NEOWISE-R data, which was obtained
significantly after the transient. The multi-wavelength SED at two
different epochs is shown in Figure \ref{fg:SED}. We find that if we
model the optical bands as an AGN, the WISE bands are significantly
overpredicted. This implies that the transient event cannot correspond
to a flare in the accretion disk, a TDE or another kind of transient
within the accretion disk, as if that had been the case, the best-fit
SED to the optical data would underpredict the emission in the WISE
bands instead due to the light travel time from the accretion disk to
the dust torus. We also find that the shape and peak luminosity of the
lightcurve are inconsistent with what would be expected for TDEs,
further ruling out that possibility.

We argue instead that the transient in \sname\ corresponds to a
SLSN. The shape of the $V_{\rm CSS}$ lightcurve and its peak
luminosity are consistent with observations of other
SLSN. Furthermore, the UV/optical multi-wavelength SED is consistent
with a blackbody with a temperature of $13,600\pm 500~\rm K$ coupled
with H$\alpha$ and H$\beta$ emission lines with large equivalent
widths of $960\pm 340~\rm \AA$ and $240\pm 90~\rm \AA$. These values
are consistent with what has been previously observed for SLSN. Using
the conservative assumption that the temperature of the SED did not
change throughout the transient, we estimate a total radiated energy
of $E=1.6\pm 0.3\times 10^{52}~\rm erg$. As the temperature of the
SLSN would have likely been much hotter closer to the peak luminosity,
the total radiated energy might be higher. We further speculate that,
because the dust temperature does not evolve throughout the transient
or between the AllWISE and NEOWISE-R epochs, the location of the SLSN
might have been within the torus of the AGN, as in that case the
accretion disk emission would maintain the dust temperature regardless
of the transient luminosity. Furthermore, the torus may be able to
provide the dense CSM that is thought to be needed to make SNe
super-luminous.

Finally, we identify 9 other highly variable, radio-undetected WISE
AGN candidates that could be analogs of \sname\ based on their mid-IR
lightcurves. However, despite many of these objects being classified
as type 1 AGN, their $V_{\rm CSS}$ lightcurves do not show significant
variability. This is further consistent with our suggestion that the
transient of \sname\ corresponds to an SLSN in the torus of the AGN,
as the majority of events could be obscured by dust in the
torus. Based on the selection function of \sname\ we estimate that the
rate of such transients is $>2\times 10^{-7}~\rm yr^{-1}$ per AGN. If
these 9 objects are true analogs to the transient in \sname, we
instead estimate a larger rate of $>7.5\times 10^{-6}~\rm yr^{-1}$ per
AGN. If we consider these events in isolation and estimate the
comoving volume in which they would have been selected by our study,
we find that their rate would be too high to be compatible with being
SLSN unrelated to the AGN activity, further suggesting a relation
between the two. However, as we cannot ascertain that these additional
objects are true analogs to the transient in \sname, our rate
estimates are not stringent enough to confirm this scenario. Future
studies relying on the much more extensive NEOWISE-R observations to
identify new candidates will be needed to further constrain the nature
of similar mid-IR transient events.

\acknowledgments

The authors would like to thank Krisztina Gabanyi and the anonymous
referee for comments and suggestions that helped improve the
article. The authors would also like to thank Cristopher S. Kochanek
for discussions and suggestions that helped making this a better
paper. We would like to thank Hannah Earnshaw, Marianne Heida, Ashish
Mahabal and Ga\"el Noirot for obtaining observations used in this
article. RJA was supported by FONDECYT grant number 1151408. Support
for JLP is provided in part by the Ministry of Economy, Development,
and Tourism’s Millennium Science Initiative through grant IC120009,
awarded to The Millennium Institute of Astrophysics, MAS. HDJ is
supported by the Basic Science Research Program through the National
Research Foundation of Korea (NRF) funded by the Ministry of Education
(NRF-2017R1A6A3A04005158). This research has made use of the NASA/IPAC
Infrared Science Archive, which is operated by the Jet Propulsion
Laboratory, California Institute of Technology, under contract with
the National Aeronautics and Space Administration. This publication
makes use of data products from the Wide-field Infrared Survey
Explorer, which is a joint project of the University of California,
Los Angeles, and the Jet Propulsion Laboratory/California Institute of
Technology, and NEOWISE, which is a project of the Jet Propulsion
Laboratory/California Institute of Technology. WISE and NEOWISE are
funded by the National Aeronautics and Space Administration. The
Pan-STARRS1 Surveys (PS1) have been made possible through
contributions of the Institute for Astronomy, the University of
Hawaii, the Pan-STARRS Project Office, the Max-Planck Society and its
participating institutes, the Max Planck Institute for Astronomy,
Heidelberg and the Max Planck Institute for Extraterrestrial Physics,
Garching, The Johns Hopkins University, Durham University, the
University of Edinburgh, Queen's University Belfast, the
Harvard-Smithsonian Center for Astrophysics, the Las Cumbres
Observatory Global Telescope Network Incorporated, the National
Central University of Taiwan, the Space Telescope Science Institute,
the National Aeronautics and Space Administration under Grant
No. NNX08AR22G issued through the Planetary Science Division of the
NASA Science Mission Directorate, the National Science Foundation
under Grant No. AST-1238877, the University of Maryland, and Eotvos
Lorand University (ELTE) and the Los Alamos National
Laboratory. Funding for SDSS-III has been provided by the Alfred
P. Sloan Foundation, the Participating Institutions, the National
Science Foundation, and the U.S. Department of Energy Office of
Science. {\tt{The SDSS-III web site is
    http://www.sdss3.org/}}. SDSS-III is managed by the Astrophysical
Research Consortium for the Participating Institutions of the SDSS-III
Collaboration including the University of Arizona, the Brazilian
Participation Group, Brookhaven National Laboratory, Carnegie Mellon
University, University of Florida, the French Participation Group, the
German Participation Group, Harvard University, the Instituto de
Astrofisica de Canarias, the Michigan State/Notre Dame/JINA
Participation Group, Johns Hopkins University, Lawrence Berkeley
National Laboratory, Max Planck Institute for Astrophysics, Max Planck
Institute for Extraterrestrial Physics, New Mexico State University,
New York University, Ohio State University, Pennsylvania State
University, University of Portsmouth, Princeton University, the
Spanish Participation Group, University of Tokyo, University of Utah,
Vanderbilt University, University of Virginia, University of
Washington, and Yale University. The CSS survey is funded by the
National Aeronautics and Space Administration under Grant
No. NNG05GF22G issued through the Science Mission Directorate
Near-Earth Objects Observations Program.  The CRTS survey is supported
by the U.S.~National Science Foundation under grants AST-0909182.

\appendix

\section{Highly Variable Radio-Loud WISE AGN}\label{app:radio_loud}

As discussed in the main text, it is likely that the majority of AGN
candidates with highly variable WISE fluxes that have a FIRST survey
counterpart are blazars. Blazars can be bright throughout the
electromagnetic spectrum, from the $\gamma$-rays to the radio. Here we
look at the WISE colors of these sources to see if they differ
significantly from those that are not detected by FIRST.

\citet{massaro12}, following the results of an earlier work,
\citet{massaro11}, developed a scheme to identify in the WISE data the
counterparts of $\gamma$-ray sources using mid-IR colors. For this,
\citet{massaro12} defined a WISE $\gamma$-ray strip (WGS) of WISE
colors on which the counterparts of $\gamma$-ray sources are typically
found. Specifically, \citet{massaro12} defined the boundaries of the
WGS in the W1--W2 vs. W2--W3, W2--W3 vs. W3--W4, and W1--W2 vs. W3--W4
color-color diagrams, independently for BL Lacs and flat spectrum
radio quasars. \citet{massaro12} furthermore defines the total strip
parameters $S_b$ and $S_q$ that quantify how consistent a given WISE
source is with being in the WGS region given its colors and errors on
those colors. A value of zero means that the source is inconsistent
with the respective WGS.

\begin{figure}
  \begin{center}
    \plotone{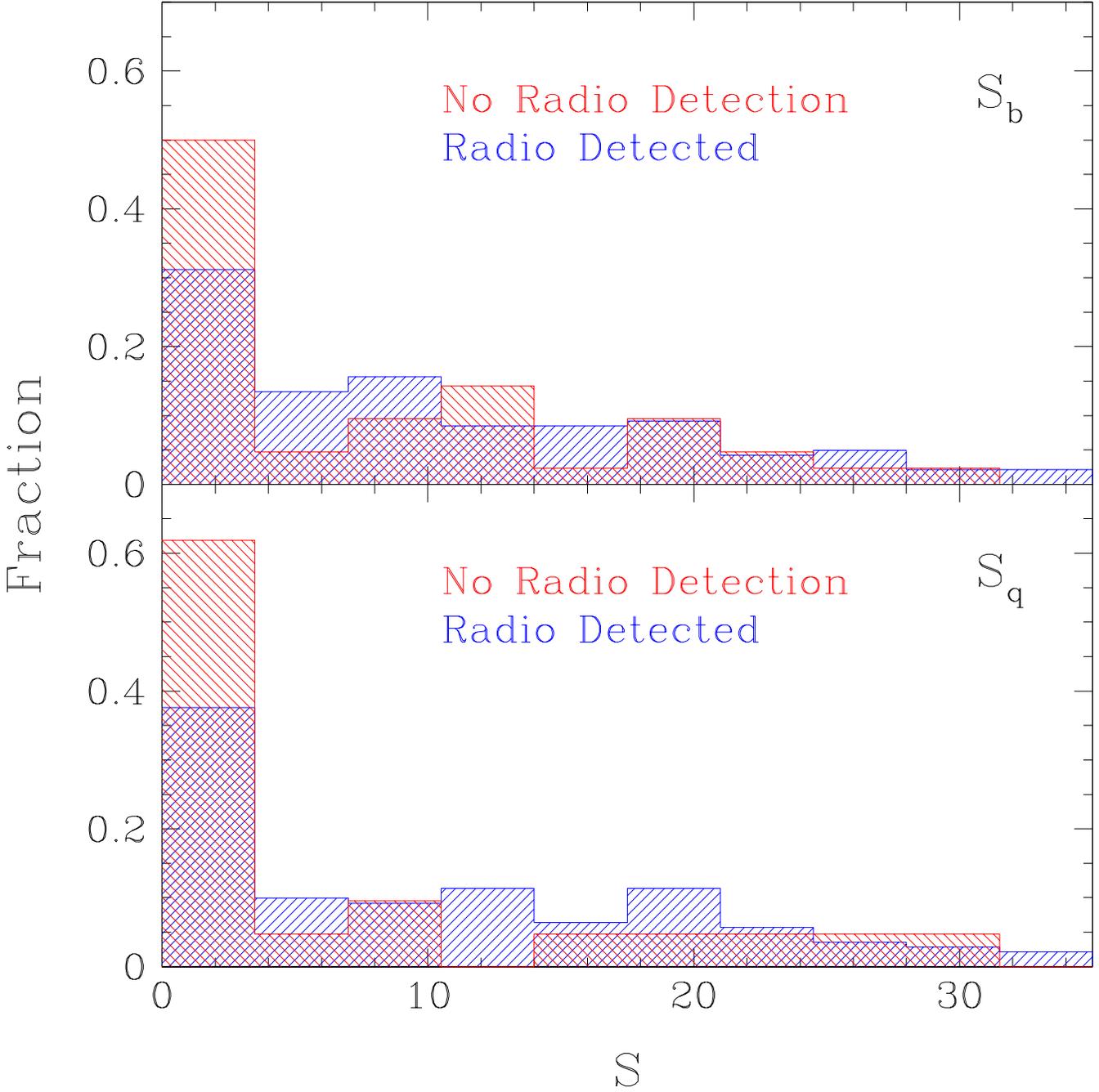}
    \caption{The distribution of the BL Lac $S_b$ (top) and flat
      spectrum radio quasar $S_q$ (bottom) parameters of
      \citet{massaro12} for the highly variable WISE AGN
      candidates. The blue histograms show these values for the
      objects that are detected in the radio by FIRST, while the red
      histograms show them for the radio-undetected highly variable
      WISE AGN candidates. Objects with $S_b=0$ or $S_q=0$ are
      inconsistent with the respective WGS. The Figure shows that the
      WGS is effective at identifying all quasars, not just BL Lacs or
      flat spectrum radio quasars.}
    \label{fg:wgs_sb_sq}
  \end{center}
\end{figure}

Our group of highly variable WISE AGN can be used to provide an
interesting test on the WGS. While the WGS was defined specifically to
aid in the identification of counterparts to $\gamma$-ray sources, we
can test whether the WGS can be used to identify blazars simply from
the WISE photometry. Specifically, we can test whether our highly
variable WISE sources with radio counterparts, likely to be blazars,
have different $S_b$ and $S_q$ parameters from those that do not have
radio counterparts, and hence are unlikely to be blazars. Figure
\ref{fg:wgs_sb_sq} shows the distribution of the $S_b$ and $S_q$
parameters for both subsamples. The objects with no-radio detections
are clearly more clustered towards $S_b=0$ and $S_q=0$. However both
samples have a large fraction of objects extending to similarly large
$S_b$ and $S_q$ values, implying that these parameters are not a good
discriminant of whether a WISE selected AGN is a blazar or not. This
is not surprising as, in addition to the WGS not being built with this
application in mind, the WGS projection in the W1--W2 vs. W2--W3 plane
overlaps strongly with the AGN selection criteria of \citet{mateos12},
which does not aim to identify blazars, i.e., the WGS is effective at
identifying all quasars, not just BL Lacs and flat spectrum radio
quasars.

\section{Optical and Mid-IR Lightcurves for the Rest of the Sample}\label{app:lightcurves_others}

Throughout the text the lightcurves of only 19 of the 45 highly
variable, radio-undetected WISE sources have been presented. Here, we
provide, for completeness, the lightcurves of the reminder of the
sample. These lightcurves are shown in Figures
\ref{fg:app_lcs_1}--\ref{fg:app_lcs_4}. Objects are arranged from left
to right, top to bottom by their $V_{\rm CSS}$-band magnitudes. As
mentioned in the text, WISEA J144039.30+612748.1 is outside the
footprint of the CRTS survey and hence no optical lightcurve is
available for it. We have put this object last.

\begin{figure}
  \begin{center}
    \plotone{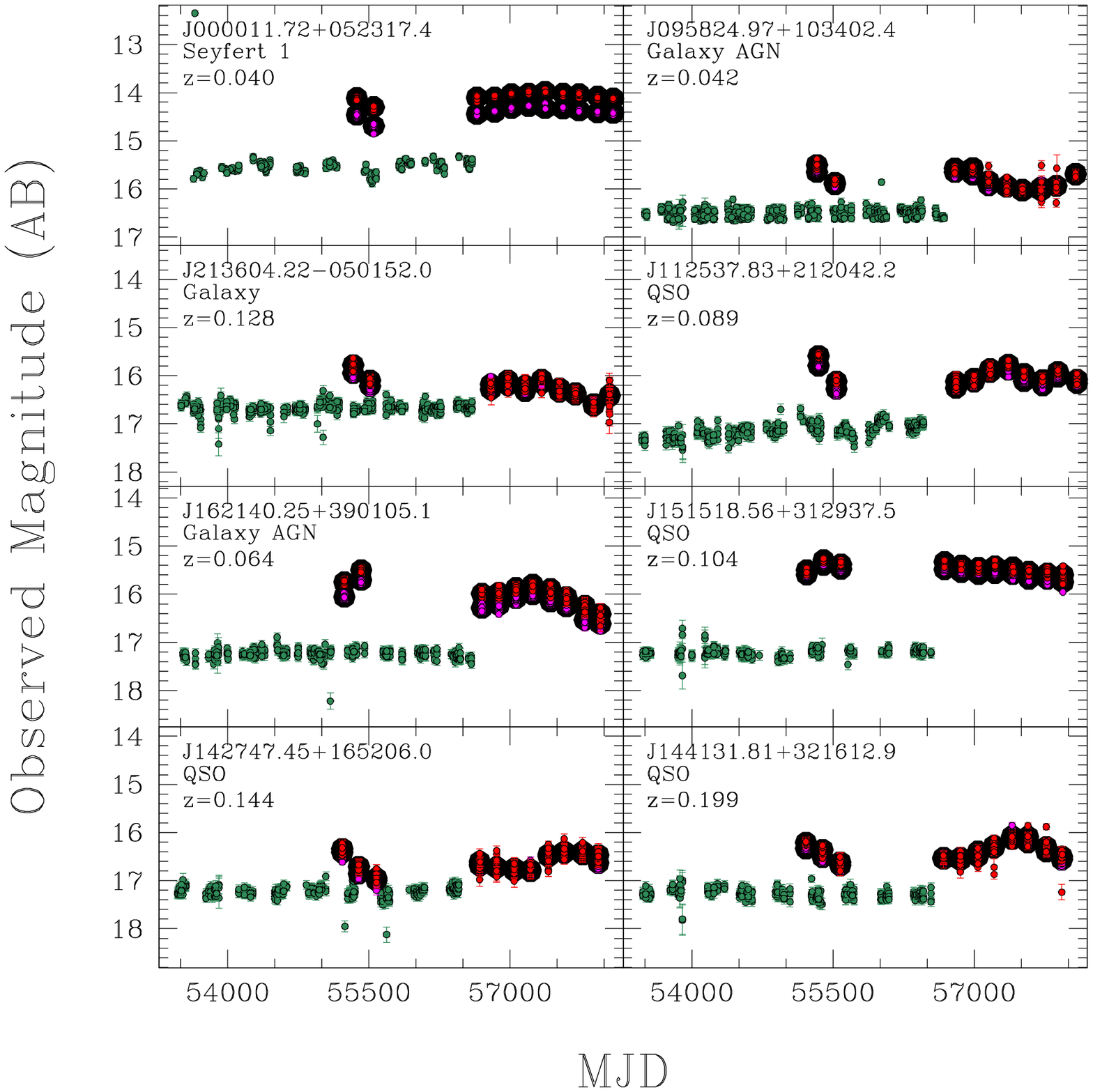}
    \caption{Lightcurves of the rest of the sample of highly variable
      WISE AGN candidates that are not detected by FIRST. Objects are
      arranged by their $V_{\rm CSS}$-band magnitude, and all
      magnitudes shown are in the AB system. Symbols and colors have
      the same meaning as in Fig. \ref{fg:lcs_high_var}.}
    \label{fg:app_lcs_1}
  \end{center}
\end{figure}

\begin{figure}
  \begin{center}
    \plotone{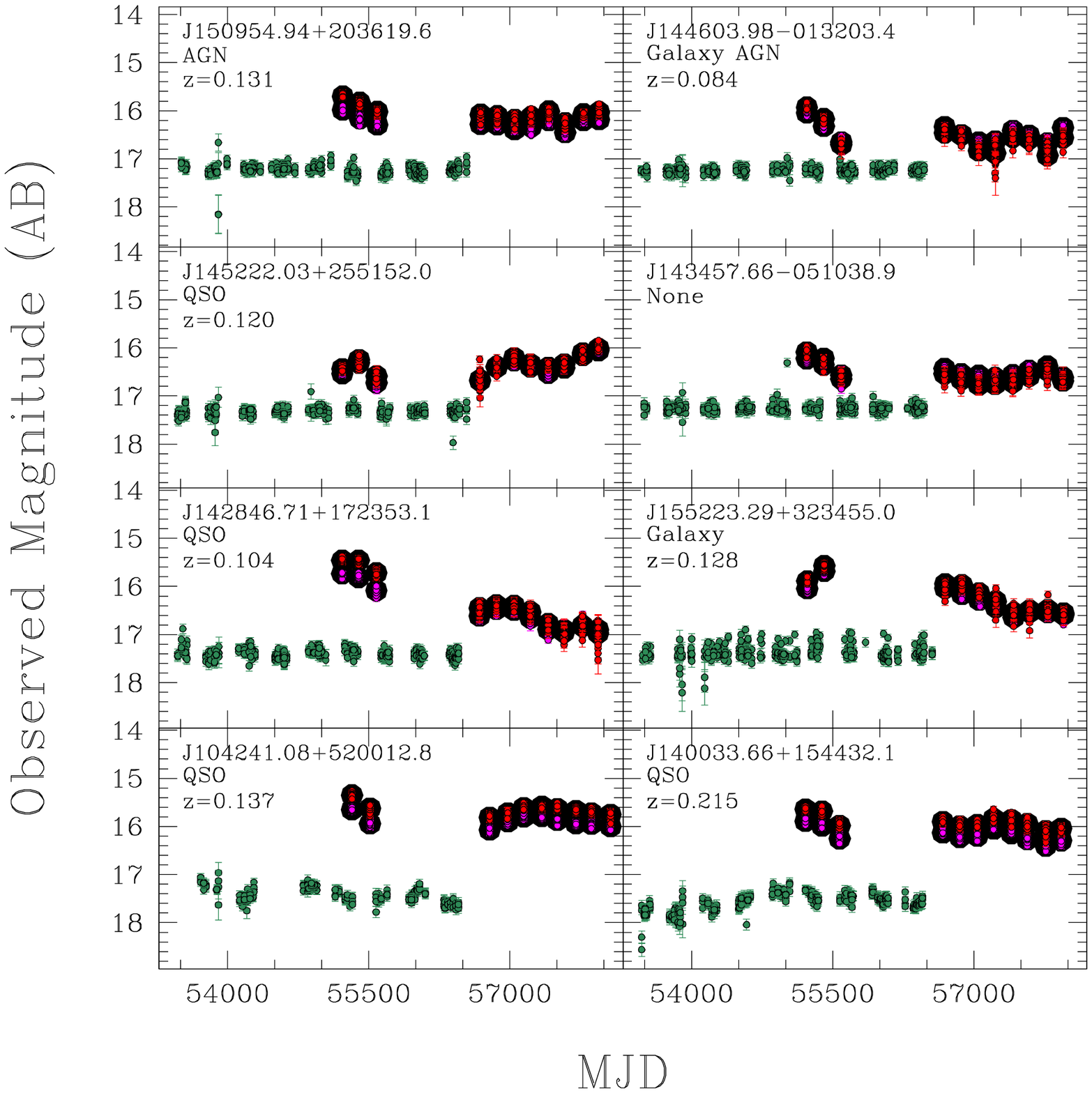}
    \caption{Continuation of Fig. \ref{fg:app_lcs_1}.}
    \label{fg:app_lcs_2}
  \end{center}
\end{figure}

\begin{figure}
  \begin{center}
    \plotone{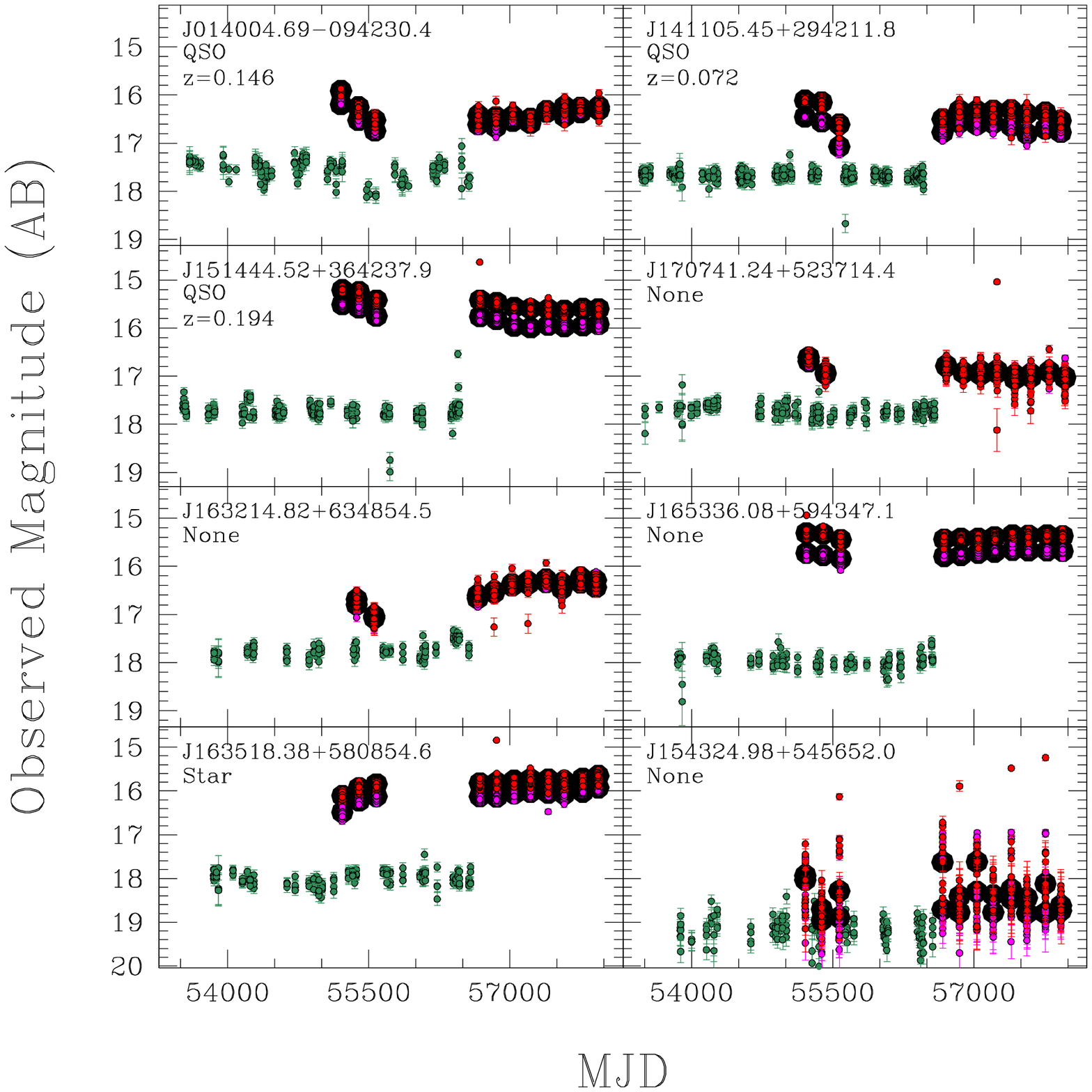}
    \caption{Continuation of Fig. \ref{fg:app_lcs_2}.}
    \label{fg:app_lcs_3}
  \end{center}
\end{figure}

\begin{figure}
  \begin{center}
    \plotone{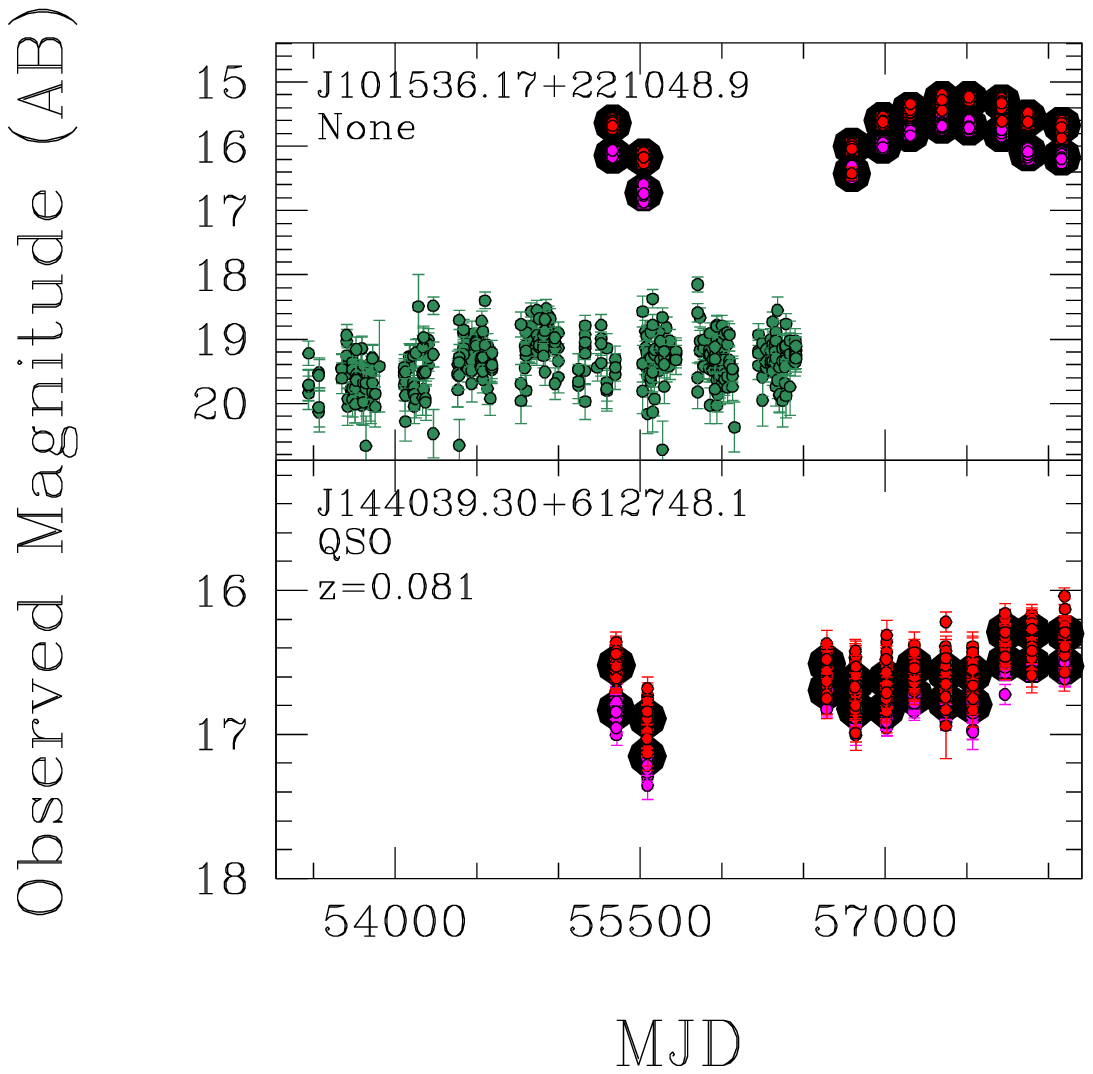}
    \caption{Continuation of Fig. \ref{fg:app_lcs_3}.}
    \label{fg:app_lcs_4}
  \end{center}
\end{figure}

\section{Optical and Mid-IR Lightcurves for the Rest of the Sample}\label{app:graham}

As discussed in \S\ref{ssec:var_agn_lightcurves}, \sname\ was also
identified by \citet{graham17} as one of 51 AGN in the CRTS database
whose lightcurves strongly deviated from a DRW model. Here we provide,
for completeness, the optical and WISE lightcurves for 49 of the other
50 objects of that sample. The missing objects, CRTS2326+0005, is not
found within the CRTS or WISE databases, suggesting a mistake in the
coordinates provided by \citet{graham17}. The CRTS and WISE
lightcurves of these 49 objects are shown in Figures 17--23.

\end{document}